\documentclass[a4paper]{article}
\usepackage[utf8]{inputenc}
\usepackage{amsmath, amsfonts, amsthm, amssymb, bbm, bm}
\usepackage{mathrsfs}
\usepackage{centernot}
\usepackage{enumerate}
\usepackage[shortlabels]{enumitem}
\usepackage{tabularx}
\usepackage{multicol}
\usepackage{multirow}
\usepackage[colorlinks=true,linktoc=all, allcolors=black]{hyperref}
\usepackage{setspace}
\usepackage[margin=3cm]{geometry}
\usepackage[round]{natbib}
\usepackage{booktabs}
\usepackage{tikz}
\usetikzlibrary{shapes,arrows,calc}
\usepackage{algorithm} 
\usepackage{algpseudocode} 
\usepackage{lscape}
\usepackage{authblk}
\usepackage[english]{babel} 
\usepackage{blindtext}

\newcommand{\E}{\mathbb{E}}

\DeclareMathOperator{\Var}{Var}
\DeclareMathOperator{\sd}{sd}

\DeclareMathOperator*{\argmax}{arg\,max}
\DeclareMathOperator*{\argmin}{arg\,min}

\DeclareMathOperator{\expit}{expit}

\DeclareMathOperator{\bernoulli}{Bernoulli}
\DeclareMathOperator{\RMSE}{RMSE}

\DeclareMathOperator{\Bias}{Bias}

\DeclareMathOperator{\var}{Var}
\DeclareMathOperator{\cov}{Cov}

\newcommand{\cmid}{\,|\,}

\newcommand\indep{\protect\mathpalette{\protect\independenT}{\perp}}
\def\independenT#1#2{\mathrel{\rlap{$#1#2$}\mkern2mu{#1#2}}}

\theoremstyle{plain}
\newtheorem{lem}{Lemma}[section]

\theoremstyle{definition}
\newtheorem{dfn}[lem]{Definition}

\newtheorem{exm}{Example}

\newcommand{\bitem}{\begin{itemize}}
\newcommand{\eitem}{\end{itemize}}

\newcommand{\barr}{\begin{array}}
\newcommand{\earr}{\end{array}}

\newcommand{\bmat}{\begin{pmatrix}}
\newcommand{\emat}{\end{pmatrix}}

\newcommand{\bs}{\boldsymbol}

\def\bal#1\eal{\begin{align*}#1\end{align*}}

\newtheorem{assumption}{Assumption}

\newcolumntype{L}{>{$}l<{$}}
\newcolumntype{C}{>{$}c<{$}}

\usepackage{tikz}
\usetikzlibrary{shapes, arrows, arrows.meta, calc, positioning}

\tikzset{nv/.style={circle, color=red, fill=red, inner sep=0.5mm}}
\tikzset{rv/.style={circle, draw, thick, minimum size=7mm, inner sep=0.5mm}}
\tikzset{fv/.style={rectangle, draw, thick, minimum size=7mm, inner sep=0.5mm}}
\tikzset{lv/.style={circle, color=red, fill=gray!30, draw, thick, minimum size=7mm, inner sep=0.5mm}}
\tikzset{rve/.style={ellipse, draw, thick, minimum size=7mm, inner sep=0.5mm}}

\tikzset{rvs/.style={circle, draw, thick, minimum size=6mm, inner sep=0.5mm}}
\tikzset{fvs/.style={rectangle, draw, thick, minimum size=6mm, inner sep=0.5mm}}
\tikzset{lvs/.style={circle, color=red, fill=gray!30, draw, thick, minimum size=6mm, inner sep=0.5mm}}
\tikzset{rves/.style={ellipse, draw, thick, minimum size=6mm, inner sep=0.5mm}}

\tikzset{deg/.style={->, very thick, color=blue}}
\tikzset{degl/.style={->, very thick, color=red}}
\tikzset{beg/.style={<->, very thick, color=red}}
\tikzset{cdeg/.style={{Circle[length=+2pt 2.5,width=+2pt 2.5, fill=none]}->, very thick, color=blue}}
\tikzset{cceg/.style={{Circle[length=+2pt 2.5,width=+2pt 2.5, fill=none]}-{Circle[length=+2pt 2.5,width=+2pt 2.5, fill=none]}, very thick}}
\tikzset{uceg/.style={{Circle[length=+2pt 2.5,width=+2pt 2.5, fill=none]}-, very thick}}
\tikzset{ueg/.style={very thick}}

\definecolor{oxblue}{RGB}{0, 33, 71}
\usepackage{booktabs}

\title{Data fusion for efficiency gain in ATE estimation:\\  A practical review with simulations}

\author[1]{Xi Lin}
\author[2]{Jens Magelund Tarp}
\author[1]{Robin J. Evans}
\affil[1]{Department of Statistics, University of Oxford, UK}
\affil[2]{Novo Nordisk, Denmark}
\date{\today}

\usepackage{makecell}

\begin{document}

\maketitle



\begin{abstract}
The integration of real-world data (RWD) and randomized controlled trials (RCT) is increasingly important for advancing causal inference in scientific research. This combination holds great promise for enhancing the efficiency of causal effect estimation, offering benefits such as reduced trial participant numbers and expedited drug access for patients. Despite the availability of numerous data fusion methods, selecting the most appropriate one for a specific research question remains challenging. This paper systematically reviews and compares these methods regarding their assumptions, limitations, and implementation complexities. Through simulations reflecting real-world scenarios, we identify a prevalent risk-reward trade-off across different methods. We investigate and interpret this trade-off, providing key insights into the strengths and weaknesses of various methods; thereby helping researchers navigate through the application of data fusion for improved causal inference.

\noindent \textbf{Keywords:} Causal inference, clinical trials, data fusion, efficiency gain, external controls
\end{abstract}


\section{Introduction}\label{sec:intro}
Randomized Controlled Trials (RCT) are widely considered to be the gold standard for establishing causality, providing high-quality evidence to inform high-stakes decisions in scientific, medical, econometric, and other research areas. 
However, RCTs face significant limitations; for example, in the context of rare diseases there may be very few sufferers or events, complicating patient recruitment, limiting statistical power, and raising data protection issues. These challenges highlight the need for a complementary approach.


Observational data, or real-world data (RWD) are increasingly available in abundance, as they are routinely collected  without designed intervention. Common examples include electronic health records (eHRs), government administrative datasets and online user behaviour data. RWD offers advantages of accessibility, large volumes, and the availability of long-term outcomes. However, the lack of randomization in treatment allocation can introduce bias due to confounding variables that are not measured.

The complementary features of RCTs and RWD have motivated the development of data fusion techniques. \cite{bareinboim2016causal} define data fusion as the integration of multiple datasets collected under heterogeneous conditions. Recognizing the value of real-world evidence, regulatory bodies such as the U.S Food and Drug Administrations (FDA) and European Medicines Agency (EMA) support its use, in particular for examining treatment efficacy for rare diseases. For example, the FDA approved drugs for Fabry disease based on both clinical trials and RWD due to the rarity of the disease \citep{FDAfabry}. Similarly, the EMA granted accelerated approval for a drug for metastatic Merkel cell carcinoma (mMCC) using evidence from a single-arm, open-label study compared with historical controls from eHR \citep{EMAmMCC}.

Efficiency gains are a primary goal of data fusion. Augmenting RCTs with RWD can enhance the statistical power of treatment effect estimates for both the overall population and specific subgroups.  While we utilize the richness of the RWD, a central problem is confounding bias due to the absence of randomization. To address this, some methods integrate only the control arm of RWD, known as external or historical controls, to mitigate bias. Common methods include test-then-pool \citep{viele2014use}, Bayesian dynamic borrowing \citep{ibrahim2000power,hobbs2012commensurate,schmidli2014robust}, and more recently, prognostic score adjustment \citep{schuler2020increasing}. 
When treatment data are available in RWD, including both arms can offer greater efficiency gains. The challenge lies in balancing variance reduction with the potential increase in confounding bias. Various methodologies seek to optimize this trade-off from different angles, including James-Stein style shrinkage estimators \citep{rosenman2023combining,green2005improved}, estimation under the general semi-parametric framework \citep{yang2020improved,li2023efficient,morucci2023double}, weighted average of treatment effect estimates from RCT and RWD \citep{Oberst2022,Chen2021}, bias-correction \citep{Kallus2018,YangDing2020,yang2020improved,cheng2021adaptive,van2024adaptive}, and Bayesian dynamic borrowing through a power likelihood \citep{lin2023many}. Section \ref{sec:litreview} provides a detailed review of these methods.

Apart from efficiency gains, data fusion can also address generalizability, transportability, and long-term impact studies. A limitation of RCTs is that they might not be representative of the target population due to eligibility criteria, which affects the validity of results. Data fusion addresses this by reweighting subgroups or extrapolating according to a target population represented within the RWD. Additionally, RWD can complement RCTs for long-term impact studies, extending the time horizon and predicting outcomes. For example, \cite{Athey2020} combined an RCT with administrative education data and concluded long-term benefits of smaller class sizes, and \citet{robbins2022data} fused a trial with national mortality registry data and showed that subsidized health insurance improves long-term mortality. This review, however, focuses primarily on the efficiency gains achieved through data fusion.

%



With numerous new data fusion methods being proposed in recent literature, it is important to understand which methods are suitable in which settings and how they compare. Although several surveys, such as \cite{Colnet2020,shi2023data,brantner2023methods}, have reviewed these methods, their discussions predominantly remain theoretical. These methods have not been thoroughly compared to one another in simulation studies reflecting realistic scenarios. Additionally, these methods have not been tested out extensively in real data analysis.

This paper aims to bridge this critical knowledge gap by presenting comprehensive simulations to evaluate various methods in terms of practicality, sensitivity to assumptions, and performance. Our study focuses on using individual-level RWD to augment an RCT, aiming for more efficient estimation of the average treatment effect (ATE) in a target population.
This represents a common use case in medical, economic, and public policy research. In a subsequent paper, building on learnings from this review, we will provide a detailed practical example, demonstrating the application of these fusion methods in a real-world context. Specifically, we will augment the PIONEER 6 medical trial with U.S. health insurance claims records to assess semaglutide’s impact on cardiovascular outcomes.

The remainder of this paper is organized as follows: Section \ref{sec:preliminaries} introduces the notations, causal estimands and common identification assumptions. Section \ref{sec:litreview} reviews the fusion methods designed to achieve efficiency gains. Section \ref{sec:sims} presents the results of our simulations, offering an in-depth analysis and interpretation of each method's performance.. Finally, Section \ref{sec:discuss} concludes with a discussion and offers concrete advice for practitioners, guiding them in selecting the most suitable method for their specific purposes.

\section{Preliminaries}\label{sec:preliminaries}
In this section, we introduce the notations used in this paper, various causal estimands, and assumptions that are commonly made for data fusion methodologies.
\subsection{Notations}
We assume two data sources: an RCT and some observational data. Let $A$ be the treatment, $\bs X$ be a vector of pre-treatment covariates, $Y$ be the outcome of interest, and $U$ represent unmeasured confounding. Let $S$ be a binary indicator of data source (1: RCT, 0: observational). Random variables are denoted by capital letters (e.g.,~$\bs X$, $Y$) and their instantiations by corresponding lowercase letters (e.g.,~$\bs X$, $Y$). Bold letters (e.g.,~$\bs X$) indicate vectors, as opposed to scalars.

The RCT data consist of i.i.d.~observations $\left \{V_i = \left(Y_i,A_i,\bs X_i \right ), S_i = 1):i \in \mathcal{R}\right \}$  with sample size $n_r$, and the observational data consist of i.i.d.~observations $\left \{V_i = \left(Y_i,A_i,\bs X_i \right ), S_i = 0):i \in \mathcal{O}\right \}$ with sample size $n_o$, where $\mathcal{R}$ and $\mathcal{O}$ are sample index sets for the two data sources. Each RCT observation can be modelled as sampled from a distribution $P_R$, and similarly $P_O$ for observational data. Typically, we would expect $n_r < n_o$ to represent a common scenario in practice where researchers want to augment a small-scale RCT with a larger observational dataset, yet this is not required.

\begin{figure}{\label{1:DAG}}
 \begin{center}
 \begin{tikzpicture}
[node distance=20mm, >=stealth, scale=3]
 \pgfsetarrows{latex-latex};
 \begin{scope}
 \node[rv]  (1)              {$X$};
 \node[rv, right of=1] (2) {$A$};
 \node[rv, right of=2] (3) {$Y$};
 \node[rv, red,above of=2] (4) {$U$};

 \draw[deg, dashed] (1) -- (2);
 \draw[deg] (2) -- (3);
 \draw[deg] (1) to [bend right] (3);
 \draw[degl , dashed] (4) to [bend right] (1);
 \draw[degl] (4) to [bend left] (3);
 \node[below of=2, xshift=0mm, yshift=0mm] {(a)};
 \end{scope}
 \begin{scope}[xshift=2cm]
 \node[rv]  (1)              {$X$};
 \node[rv, right of=1] (2) {$A$};
 \node[rv, right of=2] (3) {$Y$};
 \node[rv, red,above of=2] (4) {$U$};

 \draw[deg] (2) -- (3);
 \draw[degl] (4) -- (2);
 \draw[deg] (1) -- (2);
 \draw[deg] (1) to [bend right] (3);
 \draw[degl , dashed] (4) to [bend right] (1);
 \draw[degl] (4) to [bend left] (3);

 \node[below of=2, xshift=0mm, yshift=0mm] {(b)};
 \end{scope}
 \end{tikzpicture}
 \caption{Causal models for (a) RCT data and (b) the observational data. Dashed edges denote possible causal relationships. For example, In (a), if edge $X \to A$ exists, it means that the treatment is conditionally randomized, otherwise completely randomized.}
 \label{fig:DAG}
 \end{center}
\end{figure}
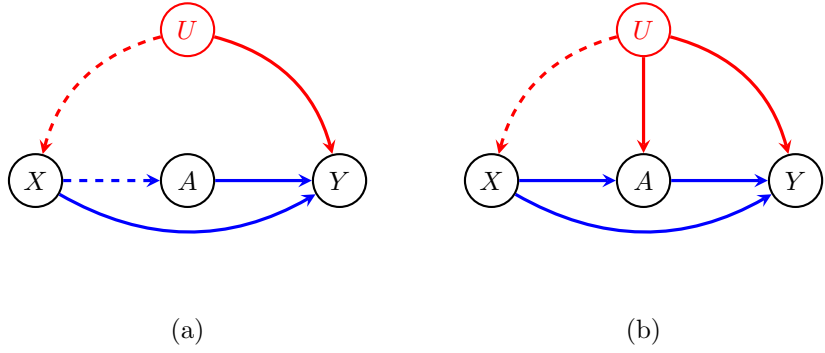

There are several overlapping frameworks to represent causal relationships, including potential outcomes \citep{rubin1974estimating}, structural causal models  \citep[e.g.][]{pearl2009causality} based on causal directed graphs \citep[e.g.][]{spirtes2000causation}, single world intervention graphs \citep{richardson2013single}, and the decision-theoretic framework \citep{dawid2021decision}. In this review, we  follow the potential outcome framework to define causal effects. Let $Y(a)$ be the potential outcome had the subject been assigned treatment $A = a$. 

\subsection{Causal estimands}
 There exists various causal quantities that researches may be interested in depending on the specific research questions. In this subsection, we introduce several of the most common causal estimands.

\begin{dfn} \label{def:ate}(Average treatment effect) $
    \tau = \E[Y(1) - Y(0)].
$
\end{dfn}
This is also known as the population average treatment effect (PATE) because it is an expectation taken across a certain target population. We can further specify exactly the population to pick out, for example, the trial population $P_R$ from which the RCT samples are drawn, the observational population $P_O$, or some other target population denoted by $P_T$.

If we further calculate the PATE conditional on subjects receiving treatment, then we derive the estimands of population average effect for the treated (PATT).
\begin{dfn}(Population average effect for the treated) $
    \tau^\text{T} = \E[Y(1) - Y(0) \mid A =1].
$
\end{dfn}
We can define the population average effect for the controls (PATC) similarly.

An alternative causal parameter is the sample average treatment effect (SATE), which is defined as the average difference in outcomes between the treated and untreated groups within a specific sample drawn from the population, rather than the population at large \citep{imbens2004nonparametric}.

\begin{dfn}(Sample average treatment effect)
$$
    \tau_S = \frac{1}{N} \sum_{i=1}^{N} [Y_i(1) - Y_i(0)].
$$
\end{dfn}
We can similarly define the sample average treatment effect for the treated (SATT) and for the control (SATC) following the same principles as the population estimands.

Another set of estimands focuses on the ATE conditional on covariate distributions and emphasizes treatment effect heterogeneity in pre-specified subgroups \citep{abadie2002simple}. 

\begin{dfn}(Conditional average treatment effect)
We define the conditional average treatment effect as:
$$
    \tau(x) = \E[Y(1) - Y(0) \cmid X = x].
$$
\end{dfn}

If the treatment effect varies by values of $X$, then we say that the treatment effect is modified by $X$, or there is treatment effect heterogeneity (HTE) across values of $X$. As a function of $X$, some papers refer to $\tau(x)$ as the HTE function\citep[e.g.][]{yang2023elastic}.

The definitions given above calculate the treatment effect as the difference in average counterfactual outcomes, identifying it as an additive effect. However, the concept of treatment effect is more flexible than this description suggests. Especially for binary outcomes, reasonable alternative estimands include odds ratios and relative risk.

\begin{dfn}(Average treatment effect as relative risk (RR) and odds ratio (OR))
\begin{align*}
    \tau^{RR} = \E Y(1) / \E Y(0), && \tau^{OR} = \frac{\E Y(1) / (1- \E Y(1) )}{\E Y(0) / (1- \E Y(0))}.
\end{align*}
\end{dfn}

In this paper, we primarily focus on the ATE defined as the difference between potential outcomes, denoted as $\tau = \E[Y(1) - Y(0)]$ in Definition \ref{def:ate}. Many methods discussed in Section \ref{sec:litreview} are developed and demonstrated using this definition. In Section \ref{sec:litreview}, we comment on how these methods might be adapted to accommodate alternative ATE estimands.

\subsection{Common identification assumptions} \label{sec:assump}
In this section, we discuss the common identification assumptions and highlight two key challenges specific to the data fusion for efficiency gains.
\begin{assumption}\label{assump:sutva} \textbf{Stable unit treatment value assumption (SUTVA) \citep{rubin1978bayesian}}

(i) No interference: Unit $i$'s potential outcomes do not depend on other units' treatments. 
$$
Y_i(a) \indep (A_1,\dots A_{i-1}, A_{i+1},\dots A_n).
$$

(ii) No multiple versions of treatment or control: the notion of being a treated (or control) subject is well-defined, so the mechanism by which this comes about does not alter the distribution of the outcome. 
\end{assumption}





    SUTVA is fundamental to the potential outcome framework, because it allows us to write $Y = Y(a)$ if $A = a$, for $a \in \{0,1\}$. This means that an individual with observed treatment $A$ equal to $a$, has observed outcome $Y$ equal to their counterfactual outcome $Y(a)$. Under the decision-theoretic framework, \cite{dawid2021decision} proposed SUTDA, which, instead of the value, assumes that the distribution of $Y_i$ only depends on the treatment $A_i$  assigned to that individual. SUTDA is a weaker condition than SUTVA and is sufficient for causal inference.

\begin{assumption}\label{assump:exchangeability} (\textbf{RCT ignorability})
(i) Exchangeability: $Y(a) \indep A  \cmid S = 1$  for $a \in \{0,1\}$, and (ii) positivity: $0 < P(A = a \cmid X, S =1) < 1 $ for all  $a \in \{0,1\}$ and $x$ in the support of the RCT.
\end{assumption}



As the first part, exchangeability assumes that the counterfactual outcomes are independent of the actual treatments, so that individuals in the treatment and control group are `exchangeable' with respect to their outcomes. A weaker version is mean exchangeability,
$\E[Y(a)\cmid A, S =1] = \E[Y(a) \cmid S = 1]$ or $\E[Y(a)\cmid A,X, S =1] = \E[Y(a) \cmid X, S = 1]$ for all $a$. The second part, positivity, requires each participant to have a non-zero probability of being assigned to treatment. Randomization is highly valued because it breaks up the confounding effect on treatment assignment, so the ignorability assumption holds for RCTs by design.



\begin{assumption} \label{assump:represent}(\textbf{RCT representativeness}) $\E[Y(1) - Y(0) \cmid S=1] = \tau$.
\end{assumption}
Assumption \ref{assump:represent} states that the RCT is representative of the target population in terms of ATE. For this review, we make this assumption to concentrate on evaluating the methods for efficiency gains without considering transportability issues---addressing one aspect at a time.

Under Assumptions \ref{assump:sutva}--\ref{assump:represent}, it is easy to see that the target causal estimand is identifiable by the RCT as $\tau = \E[Y \mid A =1, S = 1] - \E[Y \mid A =0, S = 1]$ and can be estimated through  strategies of various complexity such as difference in means, inverse propensity weighting (IPW) or augmented inverse propensity weighting (AIPW) \citep{Robins1994,robins2000robust}. These assumptions are generally acceptable due to the controlled nature of RCTs. For the target parameter to be identifiable in RWD, similar assumptions must be in place. However, these are unlikely to be met in real-world settings, presenting us with two \textbf{key challenges of data fusion in achieving efficiency gains}.

\begin{assumption}\label{assump:comparable} (\textbf{RWD comparability}) $\E[Y(a) \cmid X, S = 0] = \E[Y(a) \cmid X, S = 1]$ for $a \in \{0,1\}$, or a weaker version: $\E[Y(1) - Y(0) \cmid X, S = 0] = \E[Y(1) - Y(0) \cmid X, S = 1]$.
\end{assumption}

Assumption \ref{assump:comparable} posits that after adjusting for baseline covariates, the expected potential outcomes in the RWD are equal to those in the RCT. This assumption is reasonable when the RCT and RWD samples originate from the same super population. For example, this can be reasonably assumed to hold when a clinical trial and the RWD are collected from the same population around the same time, with no significant differences in standard of care, treatment adherence, or  measurement of outcomes and covariates. The weaker version above states that the contrasts of potential outcomes—or  causal effects—are equivalent, accommodating potential overall shifts in outcomes due to factors like time trends or systematic differences in recording outcomes.



\begin{assumption}\label{assump:rwd_exchangeability} (\textbf{RWD ignorability})
$Y(a) \indep A  \cmid X, S = 0$  and $0< P(A = a \cmid X, S = 0) <1 $ for all  $a \in \{0,1\}$ and $x$ in the support of the RCT.
\end{assumption}

Assumption \ref{assump:rwd_exchangeability} states that the treatment assignment in the RWD is ignorable and that, for individuals with the same covariates $X$ covering the support of the RCT, there is a non-zero possibility of treatment in the real world. Along with Assumption \ref{assump:comparable}, the target causal estimate is identifiable by reweighting the RWD  against the RCT, as detailed in \cite{colnet2022reweighting}. However, in reality, treatment selection is never random, and the circumstances of individuals who seek specific treatments usually differ from those who do not, resulting in selection bias. Therefore, while comparability might be assumable, claiming treatment ignorability in the RWD without additional external knowledge of the environment is impractical.

So, what makes Assumptions \ref{assump:comparable} and \ref{assump:rwd_exchangeability} \textbf{key challenges of data fusion in achieving efficiency gains}? Firstly, the assumptions of comparability and treatment exchangeability cannot be falsified using RWD alone. In the presence of an RCT, potential violations can be tested. For instance, a significant discrepancy between treatment effects estimated from the RWD and those from the RCT indicates a likely violation. This principle underlies many data fusion methods discussed in Section \ref{sec:litreview}. Secondly, even when a violation is detectable through comparison with the RCT, it remains unclear whether Assumption 4 or 5 is breached. \cite{morucci2023double} present an interesting discussion on this in their Theorem 1. 




\subsection{Identification and baseline estimator} \label{sec:baseline_estimator}

To establish a baseline for assessing the benefits of data fusion compared to using RCT data alone, we use the AIPW estimator \citep{Robins1994, rotnitzky1998semiparametric}. Under Assumptions \ref{assump:sutva}--\ref{assump:represent}, the target estimand $\tau$ is identifiable using the RCT alone:

$$
\tau = \E_{R} \left \{\frac{A \left (Y - m_1(X)\right)}{e(X)} - \frac{(1-A)\left(Y - m_0(X) \right)}{1 - e(X)} + (m_1(X) - m_0(X)) \right \}
$$

where $e(X)$ is the propensity score, i.e.,~$e(X) = \E[A = 1 \cmid X ]$, and $m_a(X)$  is the conditional outcome mean, i.e.,~$m_a(X) = \E[Y \cmid a, X]$. When Assumptions \ref{assump:comparable} and \ref{assump:rwd_exchangeability} hold, $\tau$ is also identifiable in the RWD.

 The AIPW estimator combines propensity score weighting with outcome regression to provide a doubly robust method for estimating treatment effects, ensuring consistent results even if one of the models is misspecified. We define baseline estimators: 
$$
\hat{\tau}_{r} = \frac{1}{n_r} \sum_{i\in \mathcal{R}} \left[ \frac{A_i \left (Y_i - \hat m_{1,r}(X_i)\right)}{\hat e_r(X_i)} - \frac{(1-A_i)\left(Y_i - \hat m_{0,r}(X_i) \right)}{1 - \hat e_r(X_i)} + \left (\hat m_{1,r}(X_r) - \hat m_{0,r}(X_i)\right ) \right]
$$
and 
$$
\hat{\tau}_{o} = \frac{1}{n_o} \sum_{i \in \mathcal{O}}   \left[ \frac{A_i \left (Y_i - \hat m_{1,o}(X_i)\right)}{\hat e_o(X_i)} - \frac{(1-A_i)\left(Y_i - \hat m_{0,o}(X_i) \right)}{1 - \hat e_o(X_i)} + \left( \hat m_{1,o}(X_i) - \hat m_{0,o}(X_i)\right) \right], 
$$
where $\hat e_r(X)$, $\hat e_o(X)$, $\hat \mu_{a,r}(X)$, and $\hat \mu_{a,o}(X)$ are estimated propensity score and outcome in the RCT and RWD, respectively. One attractive property of the  AIPW is that it is asymptotically normal. Under Assumptions \ref{assump:sutva}--\ref{assump:represent}, provided either of the the propensity score model or the outcome regression model is correctly specified,  $\tau = \E(\hat \tau_r)$. For the RWD, if Assumptions \ref{assump:comparable} and \ref{assump:rwd_exchangeability} hold, with the same condition, $\tau = \E(\hat \tau_o)$. However, if either Assumption \ref{assump:comparable} or \ref{assump:rwd_exchangeability} is violated, then $\hat \tau_o$ is biased. We define $\delta = \E(\hat \tau_o) - \tau$ and refer to it as the ``RWD bias'' for convenience.

\section{Existing methods}\label{sec:litreview}
 This section provides a detailed review of recent data fusion techniques aimed at efficiency gains. Lacking a single taxonomy of methods, we have organized them based on their overarching philosophy. This means looking at the big-picture strategy each method uses to bring different data sources together.

Alternative grouping criteria include whether methods make parametric assumptions about the data generative model, whether they integrate external data with both treatment and control arms or only the control arm, and whether they focus on enhancing the CATE or directly on the ATE estimation. In the latter part of this section, we present a table summarizing the various characteristics of each method.

\subsection{Test-then-pool}\label{sec:test-and-pool}
Test-then-pool style approaches start with the null hypothesis of equality between the causal estimates from multiple sources and only pool if the null hypothesis is not rejected. These approaches are commonly applied in meta-analyses or multi-site studies, where results from several studies are combined. Test-then-pool is the most basic form of dynamic borrowing \citep{viele2014use}. Typically,test-then-pool approaches require the researcher to specify the size of the hypothesis test, say $\alpha = 0.025$, which offers a way to cap the Type I error inflation introduced by the external data. Essentially, test-then-pool is an ``all or nothing" approach and a major
issue with such methods is that when the experimental data is small, the hypothesis test for any discrepancy is under-powered and rarely gets rejected, resulting in the observational data being pooled even if it is biased \citep{li2020revisit}.

The \textit{Elastic Integration} approach proposed by \cite{yang2023elastic} aims at a semi-parametric efficient estimation of the HTE function. They constructed a test statistic to simultaneously falsify Assumption \ref{assump:comparable} and \ref{assump:rwd_exchangeability}:
$$
T=\left\{n_o^{-1 / 2} \sum_{i \in \mathcal{O}} \widehat{S}_{\mathrm{O}, \widehat{\psi}_{\mathrm{E}}}\left(V_i\right)\right\}^{\mathrm{T}} \widehat{\Sigma}^{-1}\left\{n_o^{-1 / 2} \sum_{i \in \mathcal{O}} \widehat{S}_{\mathrm{O}, \widehat{\psi}_{\mathrm{E}}}\left(V_i\right)\right\}
$$ 
where $\widehat{S}_{\mathrm{O}, \widehat{\psi}_{\mathrm{E}}}\left(\cdot\right)$ is the estimated semi-parametric efficiency score (SES) for an assumed HTE function $\psi$, see Section 3.2 of \cite{yang2020improved} for a detailed definition. If Assumption \ref{assump:comparable} and \ref{assump:rwd_exchangeability} hold, then the test statistic $T$ should be close to zero and a large $T$ indicates a violation. The authors proposed to only use the RCT if $T$ exceeds a threshold $c_\gamma$, otherwise, pool the two datasets together. They proposed a data-adaptive procedure to select the threshold $c_\gamma$ to minimize MSE.

\cite{dang2022cross} proposed an\textit{ Experiment-Selector Cross-Validated TMLE} (ESCV-TMLE) algorithm which 
splits the pooled data into folds,
and employs a cross-validation approach to data-adaptively select between two experiment designs: RCT only, and pooled RCT + RWD, based on potential MSE reduction. This can be broadly described as a test-then-pool approach, though it does not involve specific hypothesis testing. According to their notation, the ‘selector’ operates as follows:
$$
s_n^{\star}=\underset{s}{\operatorname{argmin}} \frac{\hat{\sigma}_{D_{\Psi_s}^*}^2}{n}+\left(\hat{\Psi}_s^{\#}\left(P_n\right)\right)^2.
$$
Intuitively, though maybe imprecisely, $s_n^{\star}$ is the `selector' to decide whether to pool the RWD. $\frac{\hat{\sigma}_{D_{\Psi_s}^*}^2}{n}$ represents the asymptotic variance derived from the efficient influence curve, and $\left(\hat{\Psi}_s^{\#}\left(P_n\right)\right)^2$ is the estimated squared bias. Essentially, the decision to pool a certain fold of the RWD is based on optimizing the MSE.

Moreover, when an negative control outcome (NCO) is available, they innovatively proposed to include the difference in NCO into the squared-bias term to improve the estimation of bias:
$$
s_n^{\star \star}=\underset{s}{\operatorname{argmin}} \frac{\hat{\sigma}_{D_{\Psi_s}^*}^2}{n}+\left(\hat{\Psi}_s^{\#}\left(P_n\right)+\hat{\Phi}_s\left(P_n\right)\right)^2.
$$
Here, $\hat{\Phi}_s\left(P_n\right)$ represents the estimated difference in the NCO between the treatment and control groups in the pooled RWD + RCT dataset. In an RCT, or in RWD without confounding, if the NCO is valid, $\Phi_s\left(P_n\right)$ should be zero. Thus, any deviation of $\hat{\Phi}_s\left(P_n\right)$ serves as a proxy of hidden confounding.

For the inclusion of an NCO to function as intended, several conditions must be met:
\begin{enumerate}[label=(\roman*)]
    \item \textbf{U-compatibility}: The unmeasured factors confounding the treatment-outcome relationship are the same as those confounding the treatment-NCO relationship. In other words, the unmeasured confounders are fully represented by the NCO, with no additional confounders between the NCO and treatment.
    \item \textbf{Same direction}: The unmeasured factors that affect the treatment-outcome relationship should affect the treatment-NCO relationship in the same direction. Otherwise the two confounding biases may cancel each other out.
\end{enumerate}

Scale is another important consideration. Any significant difference in the scale of the outcome and the NCO can alter the balance between the two biases. For example, in a case study where the authors applied the ESCV-TMLE method to study the effect of an anti-diabetic drug \citep{dang2023case}, they used rate of fracture at the NCO, which was around 0.7\%, compared to a primary outcome rate of about 4\%. In such cases, the inclusion of an NCO might not influence the trade-off by much. 

In the simulations presented in Section \ref{sec:sims}, we examine the sensitivity of the ESCV-TMLE estimator to the use of an NCO, as well as to violation of the U-compatibility assumption. 

The advantage of the ESCV-TMLE is its flexibility to combine either both arms of the RWD or only the control arm. However, one limitation is the interpretation of the estimand. For instance, if the selector combines the RWD in 60\% of cross-validation folds, it becomes challenging to interpret the underlying population upon which the treatment effect is assessed.

\subsection{Bayesian dynamic borrowing}\label{sec:bdb}
Bayesian dynamic borrowing methods are widely applied to incorporate historical studies to construct an informative prior for the outcome $Y$ or the treatment effect $\tau$. The Bayesian framework offers a natural mechanism to discount external information that may be in conflict with the trial data. 
\cite{ibrahim2000power} introduced a general \textit{power prior} approach that raises the likelihood of historical data to a power $\eta$:
$$
\pi_{\operatorname{PP}}\left(\theta \mid D_O, \eta\right) \propto L\left(\theta \mid D_O\right)^{\eta} \pi_0\left(\theta \right),
$$
where $\theta$ represents the parameters of interest, $L(\theta \cmid \cdot)$ denotes the likelihood function, and $\pi_0$ is the initial prior, which is often chosen to be uninformative. In this setup, $\eta$ is an important hyper-parameter controlling the discounting of external evidence. It is assumed to be known and can be set by expert opinion, which can be challenging in practice. The authors extended this formulation with a hierarchical specification where the uncertainty about $\eta$ is endogenized through a prior distribution $\pi(\eta)$:
$$
\pi_{\operatorname{HPP}}\left(\theta, \eta\mid D_O\right) \propto L\left(\theta \mid D_O\right)^{\eta} \pi_0\left(\theta \right) \pi\left(\eta \right).
$$

However, \cite{duan2006evaluating} and \cite{neuenschwander2009note} both 
cautioned against the above formulation as it ignores a normalizing term, which is a function of the unknown $\eta$ and thus violates the likelihood principle. They showed that this inappropriate omission leads to counter-intuitive pooling results where the influence of the external data is close to zero even when the current and historical data are compatible. They therefore proposed the \textit{normalized power prior}:
$$
\pi_{\operatorname{NPP}}\left(\theta, \eta \mid D_O\right)=\frac{L\left(\theta \mid D_O\right)^\eta \pi(\theta) \pi(\eta)}{\int_{\Theta} L\left(\theta \mid D_O\right)^\eta \pi(\theta) \mathrm{d} \theta}.
$$
A weakness of the power prior and normalized power prior approaches is that the \textit{commensurability} of the external and current data is not explicitly parameterized. Ideally, we expect the external data is considered with more weight if it appears more compatible to the current data. To this end, \cite{hobbs2011hierarchical,hobbs2012commensurate} introduced a location commensurate power prior (LCPP), where the different parameters in the current and historical data are explicitly modeled:
$$
\pi_{\operatorname{LCPP}}\left(\theta, \eta, \omega \mid D_O\right) \propto  \int \frac{\left[L\left(\theta_o \mid D_O\right)\right]^{\eta}}{\int\left[L\left(\theta_o \mid D_0\right)\right]^{\eta}\pi(\theta) d \theta_o}  \times N\left(\theta \mid \theta_0, \frac{1}{\omega}\right) d \theta_o 
 \times \operatorname{Beta}\left(\eta \mid g(\omega), 1\right) \times \pi(\omega).
$$
Here $\theta_o$ is the parameter in the historical data and $\omega$ represents the precision of how $\theta$ is distributed around $\theta_o$ and hence parameterizes the \textit{commensurability}. $g(\cdot)$ is a positive and increasing function of $\omega$, so that $\eta$ tends to take a large value when there is strong evidence of commensurability. Extending the commensurate prior approach, \cite{schmidli2014robust} proposed a meta-analytic-predictive posterior approach to borrow information from multiple historical trials, using a mixture of standard priors.

Despite the advances, specifying a hyper-parameter or a commensurability function remains a challenge and the applications are mainly to either aggregate data or simple data generating models, for example in \cite{bonapersona2021increasing}. \cite{lin2023many} proposed a novel \textit{data-adaptive power likelihood }approach, enhancing the power prior framework in three aspects: i) introduced a data adaptive method to select the learning parameter $\eta$; ii) utilized the frugal parameterization \citep{evans2023parameterizing} to support more sophisticated causal models for individual participant-level data; and iii) adopted posterior approximation to overcome the computational challenges posed by MCMC sampling for large datasets. The posterior takes the form of
$$
q_{\operatorname{PL}}(\theta ; \eta) = q\left(\theta \mid D_O, D_R, \eta \right) \propto L\left( \theta \mid D_R\right)L\left(\theta \mid D_O\right)^{\eta} \pi_0\left(\theta \right)
$$
and $\eta$ is adaptively selected by optimizing the expected log predictive density (ELPD) across the trial population $P_R$:
$$
\eta = \argmax_{\eta \in [0,1]} \E_{P_{R}}\log p_\eta\left(V \mid D_O,D_R\right) ,
$$
where $ p_\eta\left(v \mid D_O,D_R\right) = \int_\Theta p(v \mid \theta)q_{\operatorname{PL}}(\theta ; \eta) d \theta $ is the posterior predictive distribution. Essentially, the learning rate $\eta$ is selected such that the parameter posterior predicts the trial data with the best expected accuracy. 
In addition to the three aforementioned improvements on traditional Bayesian methods, the power likelihood approach offers two other significant advantages. First, it provides the flexibility to integrate either only external control or both arms of the observational data. This flexibility is valuable in real-world applications because treatments, such as new drugs or social policies, are often not available in RWD. Second, compared to non-Bayesian methods, it automatically includes uncertainty quantification through a credible interval. However, a common limitation of Bayesian methods is the challenge in optimizing the inference of a particular parameter of interest. Specifically, the power likelihood method selects $\eta$ based on optimizing the joint posterior predictive distribution instead focusing on outcome means or the ATE. This can be a drawback compared to frequentist methods, such as targeted learning using TMLE \citep{van2006targeted}, which directly targets the estimation of the ATE.

\subsection{Weighted combination}  \label{sec:weight_comb}
Another approach to data fusion is to construct an estimator that is a weighted average of $\hat \tau_r$ and $\hat \tau_o$, and the weight is determined by optimizing a certain loss function. A common choice is MSE.


\cite{Tarima2020}  and \cite{Oberst2022} proposed such a convex combination estimator $\hat{\tau}_{\operatorname{MM}} = (1- \lambda)\hat \tau_r + \lambda \hat \tau_o$ where $\lambda$ ranges from 0 to 1 and represents the weight given to the treatment effect estimate from the observational data. $\lambda$ is chosen to minimizing the MSE. That is,
\begin{align*}
    \lambda &= \argmin_{\lambda \in [0,1]} \E\left [ \left ( \hat{\tau}_{\operatorname{MM}} - \tau_0)^2 \right)\right] \\
    & =\frac{\var(\hat{\tau}_r)-\cov(\hat{\tau}_r,\hat{\tau}_o)}{\delta^2+ \var(\hat{\tau}_r - \hat{\tau}_o)}.
\end{align*}
Since $\delta$ is unknown, the authors proposed to plug in the difference between the RCT and observational data estimates, i.e.~$\hat \delta = \hat{\tau_o} - \hat{\tau_r}$ as an estimate of the bias, to find an estimate of the MSE-minimizing weight $\hat \lambda$. \cite{Oberst2022} gave a bound on the ratio of MSE over using RCT alone in their Theorem 2, which takes a maximum of $2.25$ when $\hat \tau_r $ and $\hat \tau_o$ are independent. It is worth noting that while $\hat{\tau}_r - \hat{\tau}_o$ is an unbiased estimator of the true bias $\delta$, the plug-in estimator $\hat{\lambda}^*$ is not unbiased with regard to the MSE-minimizing weight $\lambda^*$, due to Jensen's Inequality. In fact, the plug-in estimator $\hat{\lambda}$ underestimates $\lambda^*$ on average when $\cov(\hat{\tau}_r,\hat{\tau}_o)  = 0$ due to the convexity of the expression. This means that when bias is low, their method fails to include as much influence from the observational data as is theoretically optimal. 

\cite{Chen2021} introduced a dichotomous weighted average approach with a preliminary hypothesis testing step to account for bias. The approach hinges on a comparison of bias against a pre-determined threshold. If the bias is below this threshold, the weight is selected to optimize the MSE, as though there is no bias at all; otherwise, the integration is optimized with an estimate of the bias, similar to the MSE-minimizing estimator in \cite{Oberst2022}. Their dichotomous procedure to estimate bias is as follows:
$$
\hat{\delta}_\gamma= \begin{cases}\operatorname{sgn}(\hat{\delta})\left(|\hat{\delta}|-\gamma\cdot \sqrt{\hat{\sigma}_r^2+\hat{\sigma}_o^2}\right) & \text { if }|\hat{\delta}| \geq \gamma \cdot \sqrt{\hat{\sigma}_r^2 +\hat{\sigma}_o^2} \\ 0 & \text { otherwise. }\end{cases}
$$
The selection of $\gamma$, a critical hyper-parameter, impacts the extent of bias adjustment, with a smaller $\gamma$ means that the bias term is less aggressively shrunk to zero and thus is more conservative. The authors proved that, under a certain data generating process, their estimator achieves a minimax optimal rate, up to a poly-log factor, of convergence to the oracle, where the true bias $\delta$ is unknown. However, the determination of $\gamma$'s value, which significantly influences the combination and lacks a predefined range, presents a practical challenge due to its arbitrary nature and the difficulty in justifying its selection to stakeholders.

Two significant advantages of the two described linear combination methods include: i) ease of implementation---requiring only $\hat \tau_r$, and $\hat \tau_o$ and their variances, which can be obtained from estimators like IPW. The methods only require straightforward calculations with minimal coding and does not demand extensive computational resources; ii) flexibility—the simplicity of these approaches allows for easy adaptation to alternative estimands, such as the odds ratio, by merely adjusting the definitions and estimators of $\tau_r$, and $\tau_o$. This inherent flexibility makes the method adaptable to scenarios where the outcome is binary.

Two limitations of the two methods are:  i) uncertainty quantification---their model-free nature implies that confidence intervals cannot rely on assumptions inherent in parametric models. Neither set of authors made any asymptotic claims, leading users to resort to bootstrap methods for inference; ii) applicability--by design, the two methods can only apply to integrating RWD with both treatment and control groups are available. This restricts their utility in scenarios where treatment, such as a novel drug, is not yet available in the market and absent in the RWD.

\subsection{Shrinkage} \label{sec:shrinkage}
Stein's Paradox demonstrates that, for three or more parameters, there exists a combined estimator that has a lower expected MSE than the traditional method of estimating each parameter independently, even when the independent estimators are unbiased \citep{stein1956inadmissibility,stein1961}. This result means that we gain in terms of a lower MSE by simultaneously estimating multiple parameters. The James-Stein estimator, is an application of this paradox, which shrinks the independent MLEs towards the zero vector. 

\cite{Green1991} and \cite{green2005improved} extended this idea to combining unbiased and possibly biased estimators and proposed an estimator that shrinks the unbiased estimators towards the biased estimators:
$$
\hat {\bs \tau}_{\operatorname{S1}} = \bs \delta_{2}^{+}(\hat{\boldsymbol{\tau}}_{\boldsymbol{r}}, \hat{\boldsymbol{\tau}}_{\boldsymbol{o}})=\hat{\boldsymbol{\tau}}_{\boldsymbol{o}}+
\left (\mathrm{I}_K- \frac{a \bs\Sigma_r^{-1}}{\left(\hat{\boldsymbol{\tau}}_{\boldsymbol{o}}-\hat{\boldsymbol{\tau}}_{\boldsymbol{r}}\right)^{T} \bs \Sigma_r^{-2}\left(\hat{\boldsymbol{\tau}}_{\boldsymbol{o}}-\hat{\boldsymbol{\tau}}_{\boldsymbol{r}}\right)}\right)_{+}
\left(\hat{\boldsymbol{\tau}}_{\boldsymbol{o}}-\hat{\boldsymbol{\tau}}_{\boldsymbol{r}}\right),
$$
where $\hat{\boldsymbol{\tau}}_{\boldsymbol{r}}$ and $\hat{\boldsymbol{\tau}}_{\boldsymbol{o}}$ are now vectors of dimension $K$, and each element is the ATE from one of $K$-strata in the corresponding dataset. $\Sigma_r$ is the variance-covariance matrix of $\bs \hat \tau_r$; note, that the components of $\hat{\boldsymbol{\tau}}_{\boldsymbol{r}}$ (and $\hat{\boldsymbol{\tau}}_{\boldsymbol{o}}$) are assumed mutually independent so $ \bs {\Sigma_r}$ is a diagonal matrix. The shrinkage parameter $a$ is customarily set as $K -2 $. The plus sign denotes taking the element-wise positive part of the matrix.

In the same vein, \cite{rosenman2023combining} proposed a general procedure for deriving shrinkage estimators in this setting, by minimizing Stein's Unbiased Risk Estimate (SURE). They developed a new estimator,
$$
\hat {\bs \tau}_{\operatorname{S2}} = \hat{\boldsymbol{\tau}}_{\boldsymbol{o}}+\left(\boldsymbol{I}_K-\frac{\operatorname{Tr}\left(\boldsymbol{\Sigma}_r^2 \right) \boldsymbol{\Sigma}_r}{\left(\hat{\boldsymbol{\tau}}_{\boldsymbol{o}}-\hat{\boldsymbol{\tau}}_{\boldsymbol{r}}\right)^T \boldsymbol{\Sigma_r}^2 \left(\hat{\boldsymbol{\tau}}_{\boldsymbol{o}}-\hat{\boldsymbol{\tau}}_{\boldsymbol{r}}\right)}\right)_{+}\left(\hat{\boldsymbol{\tau}}_{\boldsymbol{o}}-\hat{\boldsymbol{\tau}}_{\boldsymbol{r}}\right).
$$

Both methods require at least three strata and independence among stratum estimates, to ensure a reduction in overall MSE.  Simulations by \cite{rosenman2023combining} revealed that the choice of stratification variables and the number of strata affect performance. However, their findings remain inconclusive: it is unclear what the optimal number of strata is, whether it is better to have more or fewer strata, and no single estimator consistently outperforms the others.

For inference, both papers proposed empirical Bayes (EB) confidence intervals for the $k$th stratum-specific estimate, $\hat \tau_{\operatorname{S2},k}$, based on the result of the James-Stein estimator variance in \cite{morris1983parametric}. A key difference between EB CIs and conventional frequentist CIs is that EB CIs aim to achieve an average coverage rate for the entire $\hat {\bs \tau}$, which means that they may under-cover the causal effects for some strata and over-cover for others. Simulations in \cite{rosenman2023combining} confirmed that for large sample sizes, their 90\% EB CIs met the desired average coverage rate, although the coverage rates varied across individual strata.

Combining the weighted-combination and shrinkage concepts,  \cite{rosenman2023empirical} constructed a double shrinkage approach, which intuitively speaking, involves first constructing a MSE-minimizing weighted average of $\hat{\boldsymbol{\tau}}_{\boldsymbol{o}}$ and $\hat{\boldsymbol{\tau}}_{\boldsymbol{r}}$, and then shrinking it towards zero. To derive the estimator, the authors introduced a Bayesian hierarchical model for the `true' value of $\boldsymbol \tau$, the bias $\boldsymbol{\delta}$ and the estimates $\hat{\boldsymbol{\tau}}_{\boldsymbol{o}}$ conditional on $\boldsymbol \tau$ and $\boldsymbol{\delta}$. The expression for the $k$th entry of their double-shrinkage estimator is as follows:
\begin{align*}\lefteqn{\tau_{DS,k}\left(\phi^2, \omega^2\right)}\\
&=\underbrace{\left\{\frac{\omega^2\left(\phi^2+\sigma_{o k}^2+\sigma_{r k}^2\right)}{\sigma_{r k}^2\left(\phi^2+\sigma_{o k}^2\right)+\omega^2\left(\phi^2+\sigma_{o k}^2+\sigma_{r k}^2\right)}\right\}}_{a_k} \cdot\Bigg\{\underbrace{\frac{\left(\phi^2+\sigma_{o k}^2\right)}{\phi^2+\sigma_{o k}^2+\sigma_{r k}^2}}_{\lambda_k} \hat{\tau}_{r k}+\underbrace{\frac{\sigma_{r k}^2}{\phi^2+\sigma_{o k}^2+\sigma_{r k}^2}}_{1-\lambda_k} \hat{\tau}_{o k}\Bigg\},
\end{align*}
where $\phi$ and $\omega$ are the prior variances of $\boldsymbol \tau$ and  $\boldsymbol \delta$, respectively. The second term resembles
the form of a weighted-combination estimator, while the first term, $a_k$, acts as a shrinkage factor towards zero. Notably, $\phi$ and $\omega$ are crucial in regulating the combination, and since they are unknown, the authors suggested three estimation methods: moment matching, empirical Bayes MLE, and URE minimization. Their simulations indicated that estimating $\phi$ and $\omega$ using moment matching and EB MLE generally outperform URE minimization. Like the findings in \cite{rosenman2023combining}, simulations indicated that the performance of the double-shrinkage estimator is influenced by the selection of stratification variables and the number of strata, though the mechanism behind this influence remains unclear.

\subsection{Bias correction} \label{sec:method_bias_correct}
Another idea is to use the RCT data, which offers unbiased causal estimates, to correct the bias in the RWD estimate. The two methods in \cite{Kallus2018} and \cite{yang2020improved} are in the same vein, and the latter is a generalization of the former. By contrasting the RCT and RWD, these methods learn the impact of hidden confounding, that is, violation of Assumption \ref{assump:rwd_exchangeability}. Both of the methods assume RWD comparability (Assumption \ref{assump:comparable}) to hold.

\cite{Kallus2018} aimed at learning the impact of confounding that results in the CATE function estimated from the RWD to be biased. That is,
\begin{align*}
    \xi(X) &=\{\mathbb{E}[Y(1) \mid X, S = 0]-\mathbb{E}[Y(1) \mid X, A=1 ,S = 0]\}\\
    & -\{\mathbb{E}[Y(0) \mid X, S = 0]-\mathbb{E}[Y(0) \mid X, A=0, S =0]\},
\end{align*}

and when Assumption \ref{assump:rwd_exchangeability} holds, $\xi(X) = 0$. \cite{Kallus2018} proposed to obtain samples of $\xi(x)$  by calculating the difference between the CATE estimated in the RCT and RWD. As an illustration, they assumed a linear structure of $\xi(x)$ and derived the least-squares coefficient estimates. The resulting $\hat \xi(x)$ serves as the bias-correction term in this procedure termed \textit{experimental grounding}. They demonstrated that under linearity assumption, the bias-corrected CATE estimate $\hat \tau_{\operatorname{XG}}(x)$ is root-$n$ consistent, although efficiency gains are not theoretically proven.

\cite{yang2020improved} summarized the impact of unmeasured confounding on the potential outcomes into the \textit{confounding function}:
$$
\varphi(X)=\mathrm{E}[Y(0) \mid A=1, X, S=0]-\mathrm{E}[Y(0) \mid A=0, X, S=0].
$$

They demonstrated that, assuming the parametric forms of the confounding function $\varphi(X)$ and the CATE function $\tau(X)$ are known, both $\varphi(X)$ and $\tau(X)$ are non-parametrically identifiable. They also derived semi-parametric efficient, doubly robust estimators under several examples of assumed parametric form of $\tau(X)$. In their Theorem 3, the authors proved efficiency gains in estimating $\tau(x)$ compared to using the RCT alone, provided the terms in parametric forms of $\tau(x)$ and $\varphi(x)$ are not co-linear. However, if the terms of $\tau(x)$ are a linear combination of those in $\varphi(x)$, their is no efficiency gain from data fusion. This is empirically observed in our simulations. The authors provided some intuition for this result: when the terms are linearly dependent, all the observational data is consumed to estimate the coefficients of $\varphi(X)$ and hence does not contribute to the estimation of $\tau(x)$. While their method is demonstrated using the additive treatment effect as defined in Definition \ref{def:ate}, adapting it to binary outcomes, or alternative estimands s such as relative risks or odds ratios may present additional complexities.

While these two methods do not require RWD ignorability (Assumption \ref{assump:rwd_exchangeability}), these methods make assumptions about the parametric structure of the functional forms of CATE and confounding, which are subject to misspecification.

\cite{van2024adaptive} proposed a bias-correction method that targets a slightly different estimand: the average treatment effect averaged over the covariate distribution of the combined RCT-RWD data. They decomposed the the the ATE estimand into the difference between a pooled-ATE estimand and a bias estimand that accounts for the conditional effect of RCT enrollment on the outcome. To estimate them, they introduced an adaptive targeted minimum loss-based estimation (A-TMLE) framework. Unlike the other two bias-correction methods discussed, the A-TMLE estimator does not rely on parametric models for the biases but instead employs flexible methods for estimation.

\subsection{Prognostic adjustment}\label{sec:procova}
The EMA recently approved the historical borrowing method, PROCOVA, by \cite{schuler2020increasing} \citep{EMAprocova}. The method enhances the control arm of RCTs by incorporating a prognostic score fitted using a historical dataset, which is required to originate from the same population as the trial control arm, to ensure that mean control outcomes are the same---this is the stronger version of Assumption \ref{assump:comparable}. Although the authors suggested that this assumption can be weakened, the anticipated theoretical efficiency gain  relies on the control outcome estimated from the historical data being consistent. This may not hold if the historical RWD are sourced from different trials or countries. In addition, if outcomes are linearly related to covariates, including linearity after transformation such as in generalized linear models (GLMs), no efficiency benefit is achievable beyond that from the routine adjustment by pre-treatment covariates. Coincidentally, this resembles the intuition of the results in \cite{yang2020improved} as discussed in Section \ref{sec:method_bias_correct}; this is because all the information from the raw covariates is already exploitable by the linear model. 

The key advantage of PROCOVA is protection against Type I error, which is crucial for regulatory decisions. The authors provide a sample size calculation method based on the correlation between the prognostic score and outcome residuals after adjusted for pre-treatment covariates. Despite relying on assumptions such as equal correlation between the outcome and the prognostic score, and equal variance of the outcome between the RWD and the RCT, the ability to perform sample size calculation enables the evaluation of potential efficiency gains, and thus the reduction in required samples size, before conducting an RCT. The dual advantages of safeguarding against Type I error and enabling sample size estimation distinguish PROCOVA as the only data fusion method endorsed by regulatory agencies to date.

\subsection{A summary table for application}
Table \ref{tab:summ} provides a practical summary of the methods discussed above, serving as a reference for researchers selecting an appropriate method. The ``data assumption" column indicates whether a method is parametric (P), non-parametric (NP) or semi-parametric (SP). Understanding these assumptions helps users recognize their limitations: semi-parametric methods require correct model specification of the parametric part to avoid biased estimates, while non-parametric methods offer flexibility but may have poor finite sample performance. The next two columns indicate whether a method is applicable to combining only the control arm or both arms from the RWD. The ``inference" column describes how each approach quantifies uncertainty. The following column assesses the difficulty of implementation, specifically coding difficulty. ``Easy" denotes straightforward coding in a couple of lines, while methods like the power likelihood approach are marked as ``medium" due to the need for users to specify a causal model and assemble basic functions despite the available package. The R packages available are: \texttt{causl} \citep{causl} and \texttt{ManyData} \citep{ManyData} for the power likelihood approach, \texttt{ElasticIntegrative}\citep{ElasticIntegrative2023} for the elastic integrative estimator, \texttt{EScvtmle} \citep{ESCVTMLE2023} for the ESCV-TMLE estimator, and \texttt{IntegrativeHTEcf} \citep{HTEcf2020} for the confounding function method. See Appendix \ref{sec:implementation} for details. The ``alternative estimands" column compares the complexity of adapting methods to estimands such as relative risk. Semi-parametric methods typically require significant modification, such as replacing the efficient influence function, and current packages often lack this flexibility. The final column lists primary limitations for users to consider.

\begin{landscape}

\begin{table}[ht]  \label{tab:summ}
\small
\centering
\begin{tabular}{|p{2.5cm}|p{3.5cm}|p{1.5cm}|p{1.2cm}|p{1cm}|p{2.5cm}|p{2.5cm}|p{2cm}|p{4cm}|}
\toprule
\textbf{Category}       & \textbf{Method}         & \textbf{Data assumption} & \textbf{RWD control-arm only} & \textbf{RWD both arms} & \textbf{Inference}                    & \makecell{\textbf{Implement.}\\\textbf{difficulty}} & \textbf{Alt.~estmnds} & \textbf{Limitations/requirements}          \\ \midrule
Bayesian dynamic borrowing & \makecell{Power likelihood\\ \citep{lin2023many} } & P                       & \checkmark                      & \checkmark                &                 Bayesian credible interval         & Medium                          & Easy                           & Requires full parameterization of data distribution and correct HTE specification \\
Test-then-pool            & \makecell{Elastic integration\\ \citep{yang2023elastic}}  & NP                      &                      & \checkmark                &                 Asymptotics/ bootstrap               & Package ready                   & Hard                           & Requires correct HTE specification        \\
                           & \makecell{ESCV-TMLE\\ \citep{dang2022cross}}            & SP                      & \checkmark                      & \checkmark                &                 Asymptotics                         & Package ready                   & Hard                           & Interpretability is challenging           \\
Weighted combination       & Anchored thresholding \citep{Chen2021} & NP                      &                       &                           &                Bootstrap                          & Easy                            & Easy                           & No reversion to RCT-only with increasing bias \\
                           & MSE-minimizing \citep{Oberst2022}     & NP                      &                       & \checkmark                &                 Bootstrap                          & Easy                            & Easy                           &                                     \\
Bias correction            & \makecell{Experiment grounding\\\citep{Kallus2018}} & SP                      &                      & \checkmark                &                 bootstrap                          & Easy                            & Hard                           & Assumes linearity; Efficiency gains in HTE but not in ATE estimation  \\
      & \makecell{Confounding function\\\citep{yang2020improved}} & SP                      &                       & \checkmark                &                Asymptotics/ bootstrap               & Package ready                   & Hard                           & Requires correct HTE specification; Efficiency gains in HTE but not in ATE estimation \\
  Covariate adjustment                          & \makecell{PROCOVA\\\citep{schuler2020increasing}}            & P                       & \checkmark                      &                 &                 Sandwich estimator                  & Easy                            & Easy                           & Limited efficiency gain when outcome is nearly linear \\
Shrinkage                  & \cite{green2005improved}                       & NP                      &                       & \checkmark                &                 Bootstrap                          & Easy                            & Easy                           & Requires multiple independent strata estimates \\
                           & \cite{rosenman2023combining}                       & NP                      &                       & \checkmark                &                 Bootstrap                          & Easy                            & Easy                           & Requires multiple independent strata estimates \\ \bottomrule
\end{tabular}%
\caption{Comparison of different statistical methods.}
\end{table}
\end{landscape}

\section{Simulation studies}\label{sec:sims}
In our simulation, we aim to be as realistic as possible by introducing both binary and continuous covariates, explicitly specifying hidden confounding, and accounting for treatment heterogeneity, among other factors. Details of method implementation can be found in Appendix \ref{sec:implementation}.



\subsection{Simulation setup} \label{sec:sim_setup}

We adopt the \textit{frugal parameterization} proposed in \cite{evans2023parameterizing} to parameterize and simulate from specified marginal structural models (MSM). The primary advantage is its flexibility: we can simulate exactly the causal effects we would like. We first generate three pre-treatment covariates $X_1$, $X_2$ and $X_3$, where $X_3$ acts as an effect modifier to introduce treatment effect heterogeneity. We assume that the marginal distribution of $X_1$ is a standard normal distribution, and the marginal distribution of $X_2$ follows a Bernoulli distribution with a probability of 0.5. We model the dependency between $X_1$ and $X_2$ later using a copula. We assume that the effect modifier $X_3$ is independent of $X_1$ and $X_2$ and is normally distributed.
\begin{align*}
    X_1 &\sim \mathcal{N}(0,1)\\
    X_2 &\sim \bernoulli(0.5)\\
    X_3 &\sim \mathcal{N}(0,1)\\
    U &\sim \mathcal{N}(0, 1.1)\\
    Y(a) \mid X_3,U &\sim \mathcal{N}(0.5 + 0.2a + 0.1X_3 a + X_3 + U,1)
\end{align*}
Given the mixture of continuous and discrete variables, we model the dependence between $X_1$, $X_2$ and $Y$ using a latent Gaussian copula model, as proposed by \cite{fan2017high}. Specifically, we employ use a trivariate Gaussian copula with correlation coefficients $\rho_{X_1 X_2} = 2 \expit(0.1) -1 \approx 0.05$, $\rho_{X_1 Y} = 2 \expit(2) -1 \approx 0.76$ and $\rho_{X_2 Y} = 2 \expit(1) -1 \approx 0.46$. These coefficients allow us to simulate a weak correlation between $X_1$ and $X_2$, a strong influence of the continuous variable $X_1$ on the outcome and a moderate correlation between the binary variable $X_2$ and $Y$. 

We simulate treatment assignments in the RCT and in the observational data as below:
\begin{align*}
A_{obs} &\sim \bernoulli(\expit\{-0.5 + X_1 + X_2 + X_3 + \psi U\})\\
A_{RCT} &\sim \bernoulli(0.5).
\end{align*}
This parameterization assumes that treatment assignment is completely randomized in the RCT. In this setup, the value of the target estimand, $\tau = \E[Y(1) - Y(0)]$ is $0.25$. In contrast, in the observational data, treatment assignment depends on observed covariates $X_1$, $X_2$ and $X_3$ as well as confounding from $U$. The degree of hidden confounding is controlled by the value of $\psi$. When $\psi = 0$, $U$ is no longer a confounder, making the target estimand identifiable in the observational data. However, as the absolute value of $\psi$ increases, the bias in the ATE estimated from the RWD correspondingly intensifies. In the simulation, we vary the value of $\psi$ to introduce different magnitudes of hidden confounding into the observational data. Note that the average proportion of treated individuals remains at 50\% in all simulations.
%
%
%
We set the size of the RCT and observational datasets to 300 and 1200 respectively.

\subsection{Augmenting both arms of the RWD} \label{sec:both_arms}
\begin{figure}
    \centering
    \includegraphics[width=1\linewidth]{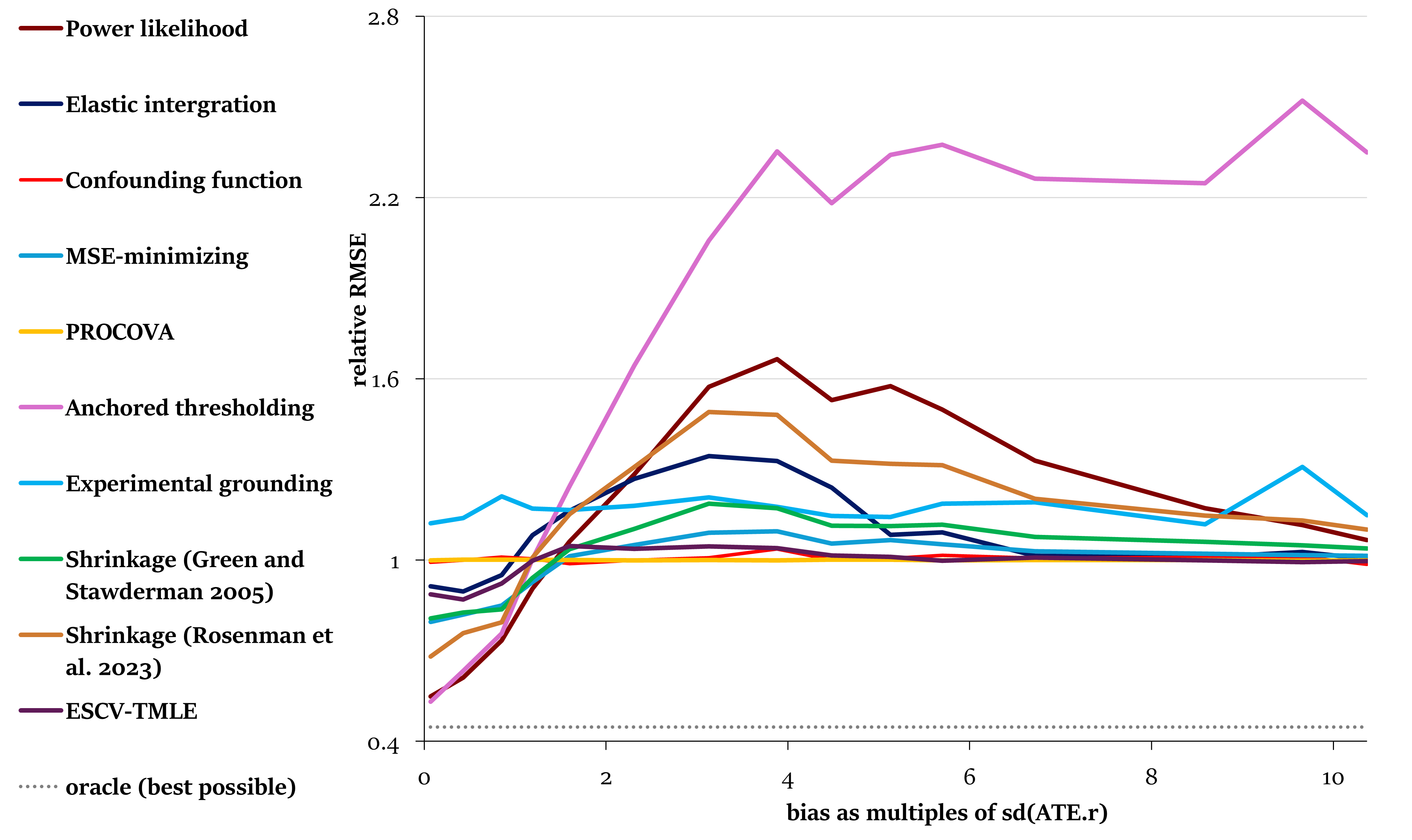}
    \caption{RMSE relative to RCT-only approach (without data fusion) by varied magnitude of bias induced by hidden confounding, in the baseline scenario with a continuous outcome}
    \label{fig:cont_base1}
\end{figure}

In Figure \ref{fig:cont_base1}, each line represents one data fusion estimator discussed in Section \ref{sec:litreview}, and the results are averaged over 300 different simulated datasets. Comparing the root mean square error (RMSE) of the fusion estimator $\hat \tau$ to that of $\hat \tau_r$, the estimator from the RCT alone, is particularly useful to determine the benefits of data fusion. The y-axis of the figure shows the relative RMSE, which is defined similarly to \cite{Oberst2022}:
\begin{align*}
    r\RMSE = \frac{\RMSE(\hat \tau)}{\RMSE(\hat \tau_{r})},
\end{align*}
where as a reference estimator, $\hat \tau_{r}$ is the AIPW estimator using the RCT data only. The fusion estimator $\hat \tau$ outperforms $\tau_{r}$ if rRMSE is less than 1 and worse, if rRMSE exceeds 1, which means that the efficiency gain is not enough to offset the inflation in bias. 

As a consequence of violation of Assumption \ref{assump:comparable} and \ref{assump:rwd_exchangeability}, the RWD bias $\delta$---defined in Section \ref{sec:baseline_estimator}---is plotted on the x-axis. Instead of the absolute value of $\delta$, we plot bias on a relative scale. Specifically, we define the relative bias as the multiple of the standard deviation of the RCT-only estimator $\hat \tau_r$:
\begin{align*}
    r\Bias = \frac{\delta}{\sd(\hat \tau_r)}.
\end{align*}
For instance, a relative bias of one indicates that $\tau_o$ is one standard deviation away from $\tau$. The advantage of using relative bias is that it provides an intuitive metric of confounding bias in relation to the variability of $\hat \tau_r$.
 
We also include an oracle performance curve for reference, which illustrates the best possible reduction in RMSE that could be achieved if the bias of the estimator $\hat\tau_o$ were known when fusing the two datasets. An example is when $r \Bias = 0$, indicating that the observational data should be integrated entirely and treated equally as the RCT data. In this scenario, the reduction in RMSE is due to the increase in total sample size, which is approximately equal to $\sqrt{300/(1200 + 300)} \approx 0.44$ in our simulation setting. We include this curve to demonstrate how close each method can come to the optimal performance. However, note that the oracle curve does not apply to the PROCOVA estimate as it only incorporates the control arm.

\begin{figure}
    \centering
    \includegraphics[width=1\linewidth]{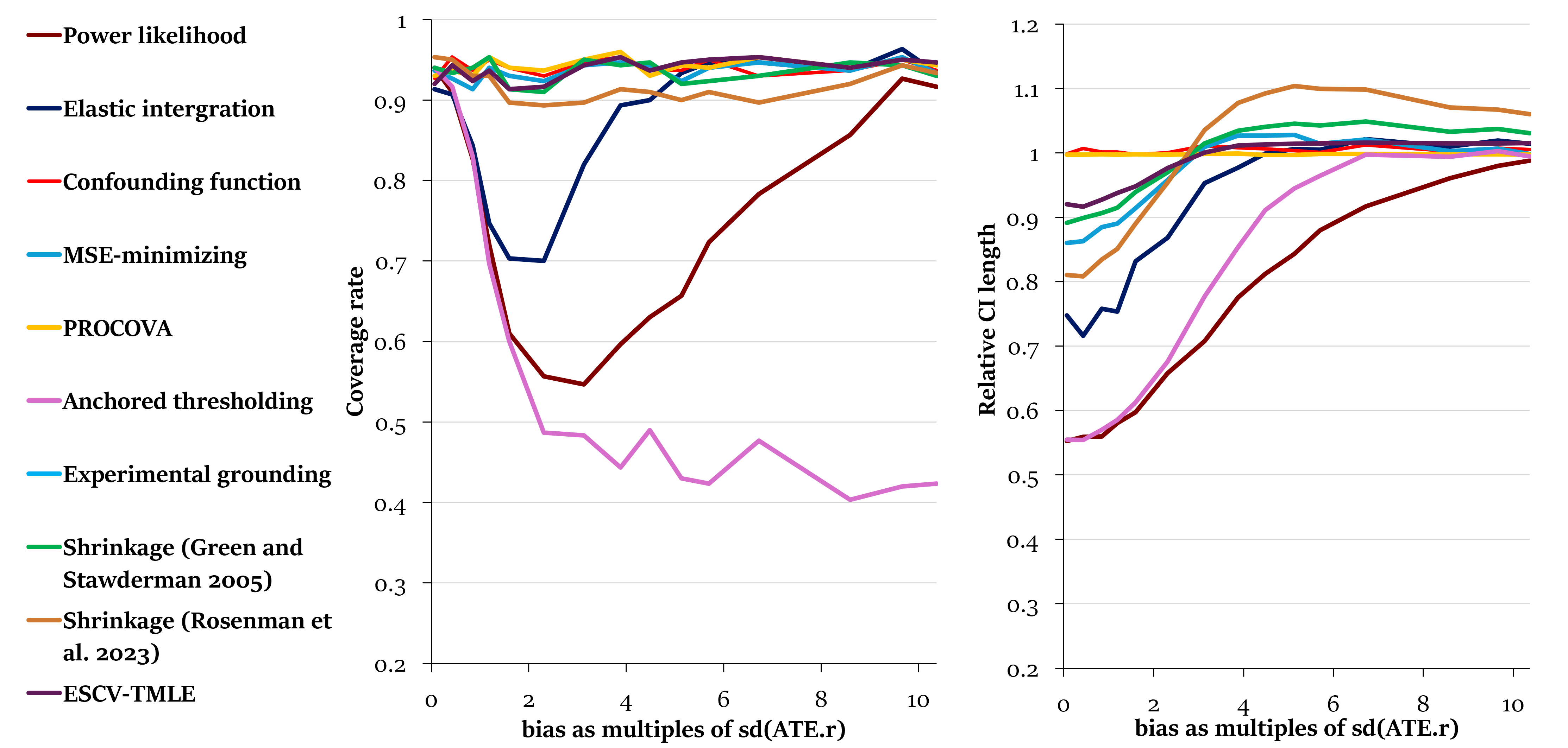}
    \caption{Coverage rate and relative CI length of data fusion estimators, by varied magnitude of bias induced by hidden confounding.}
    \label{fig:cont_cov1}
\end{figure}

Figure \ref{fig:cont_cov1} compares the coverage rates of 95\% confidence intervals (credible interval for the power likelihood estimator), and the ratio of the length of the 95\% CI of each data fusion estimator against that of $\hat \tau_r$. 
This comparison illustrates a trade-off: estimators that achieve greater power gains when bias is small are also at risk of poorer worst-case coverage rates. This highlights the trade-off between power gains and coverage rate.

\subsubsection{Anchored thresholding}

As $\delta$ increases, almost all the curves converge to the reference line, except for the anchored thresholding estimator in \cite{Chen2021}. 
The reason is that as the RWD bias, $\delta$, approaches infinity, their proposed estimator becomes asymptotically biased. We illustrate this using the following example.

\begin{exm}
    Let $\hat \tau_r \sim N(\tau, \sqrt{\sigma^2/n_e})$ and $\hat \tau_o \sim N(\tau + \delta, \sqrt{\sigma^2/n_o})$. We assume that the total sample size is $n = n_r + n_o$ and $n_o = \rho n$, where $\rho \in (0,1)$.
    
    Following the methodology outlined in Section 3.2 of \cite{Chen2021}, the weighting factor becomes $\omega = \frac{\sigma^2/n_r}{\sigma^2/n_r + \sigma^2/n_o} = \rho$, and the estimator of bias $\hat \delta = \hat \tau_o - \hat \tau_r \sim N \left(\delta, \sqrt{\frac{\sigma^2}{n_r} + \frac{\sigma^2}{n_o}}\right)$.
    We assume $\delta$ is a very large positive number, specifically, $\delta \gg \sqrt{\frac{\sigma^2}{n_r} + \frac{\sigma^2}{n_o}}$. After the proposed soft-thresholding procedure, the adjustment for bias they proposed is $\hat \delta_\gamma = \hat \delta - \gamma \sqrt{\frac{\sigma^2}{n_r} + \frac{\sigma^2}{n_o}} \sim N\left(\delta -\gamma \sqrt{\frac{\sigma^2}{n_r} + \frac{\sigma^2}{n_o}}, \sqrt{\frac{\sigma^2}{n_r} + \frac{\sigma^2}{n_o}} \right)$. Recall that $\gamma$ is an arbitrary positive hyper-parameter that controls the adjustment. Therefore, their final estimator becomes: 
\begin{align*}
    \hat \tau_{\operatorname{AT}} &= \omega \hat \tau_r + \rho \left (\hat \tau_o  - \hat \delta_\gamma \right)\\
              &=  (1-\rho)\hat \tau_r + \rho \left (\hat \tau_r +  \hat \delta - \hat \delta + \gamma \sqrt{\frac{\sigma^2}{n_r} + \frac{\sigma^2}{n_o}} \right)\\
              &= \hat \tau_r + \rho \gamma \sqrt{\frac{\sigma^2}{n_r} + \frac{\sigma^2}{n_o}}\\
              & \sim N \left(\tau + \rho \gamma \sqrt{\frac{\sigma^2}{n_r} + \frac{\sigma^2}{n_o}}, \sigma/\sqrt{n_r} \right ).
\end{align*}
\end{exm}

It is obvious that when $\delta$ is very large, their anchored-thresholding estimator $\hat \tau_{AT}$ is biased by $\rho \gamma \sqrt{\frac{\sigma^2}{n_r} + \frac{\sigma^2}{n_o}}$, with a variance equal to using the RCT data alone. Additionally, a larger $\gamma$ leads to greater bias, which explains the empirical observations made in \cite{Oberst2022}. This extra bias is the reason why the RMSE curve remains constantly above the reference line, and the coverage rate remains low as $\delta$ increases. This means that the anchored thresholding estimator does not offer the assurance that when the RWD is highly biased, we will not be worse off than using the RCT alone. For this reason, despite showing the highest power gains in Figure \ref{fig:cont_cov1}, we do not recommend the anchored thresholding method for data fusion.

\subsubsection{Bias-correction methods}

\begin{figure}
    \centering
    \includegraphics[width=1\linewidth]{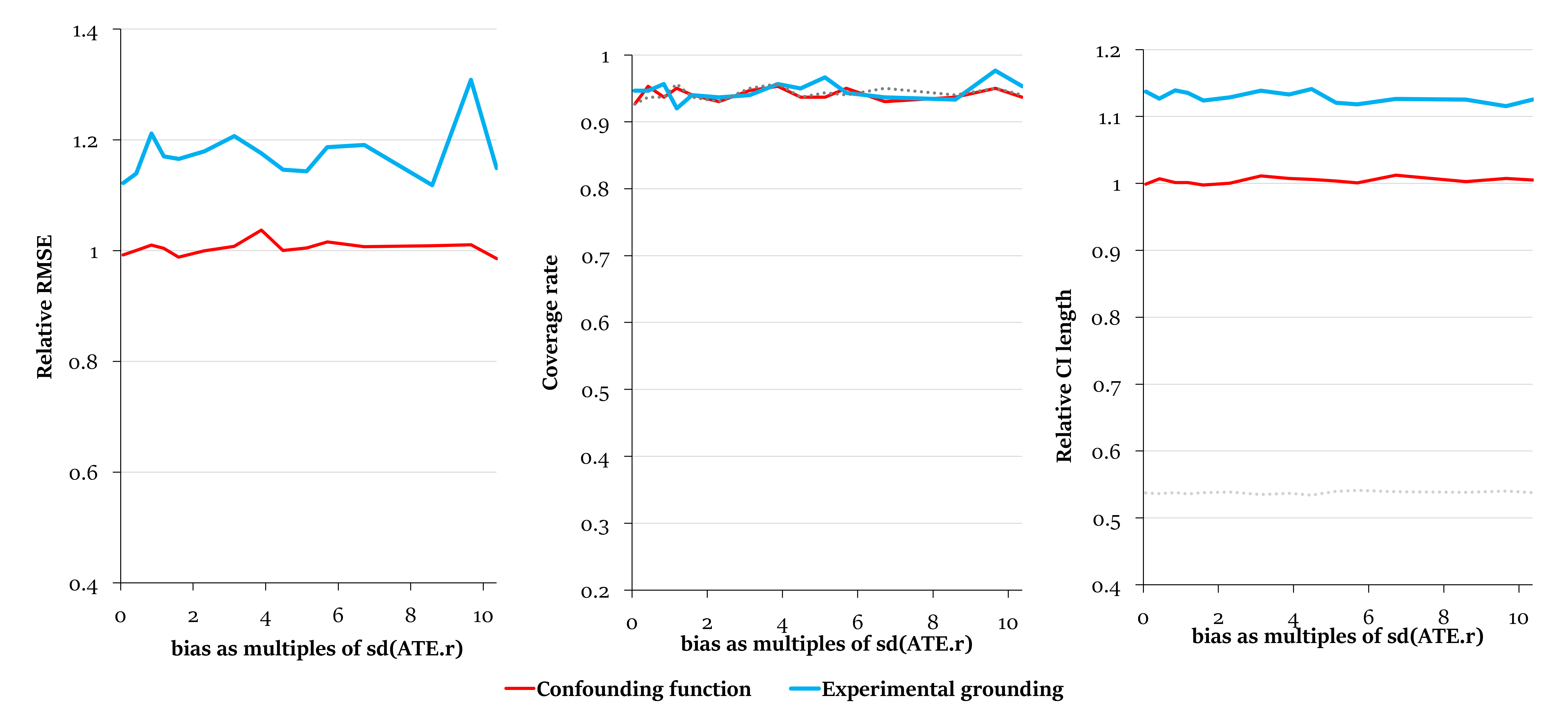}
    \caption{Relative RMSE, coverage rate and relative CI length of bias-correction estimators \citep{Kallus2018,yang2020improved}.}
    \label{fig:cont_bias_corr}
\end{figure}

Figure \ref{fig:cont_bias_corr} focuses on the two bias-corrections methods. It shows that neither the confounding function \citep{yang2020improved} nor the experimental grounding \citep{Kallus2018} reduces the RMSE, regardless of the level of confounding bias. In fact, using the experimental grounding estimator results in a loss of efficiency, leading to wider confidence intervals.

This is because both methods essentially distribute the difference between $\hat \tau_r$ and $\hat \tau_o$ across a pre-specified function space defined by the covariates. 
When marginalized over the RCT samples, the estimate reverts to $\hat \tau_r$, yielding no efficiency gains in estimating the ATE. \cite{van2024adaptive} further formalized this point by comparing the non-parametric efficiency bounds, showing that if the target parameter measures the HTE in the RCT averaged over the covariate distribution of the RCT, their may still be no efficiency gains, even if the estimand is efficiently estimated. Additionally, for the experiment grounding estimator, the loss is due to the added noise, which makes the estimation less efficient. 

It is worth noting that the confounding function method is designed for efficiency gains in estimating the HTE function, with theoretical proof provided in Theorem 3 in \cite{yang2020improved}. However, this method is not suitable for estimating the ATE in our context. If we make an extra assumption that the observational sample is representative of the target population, meaning both the RCT and RWD comes from the same population, we can marginalize the estimated HTE function over the pooled RCT + RWD data to achieve power gains by effectively increasing the sample size. Even then, the efficiency gain relies on correct specification of the function spaces.

\subsubsection{Methods showing a trade-off} \label{sec:sims_tradoff}

\begin{figure}
    \centering
    \includegraphics[width=1\linewidth]{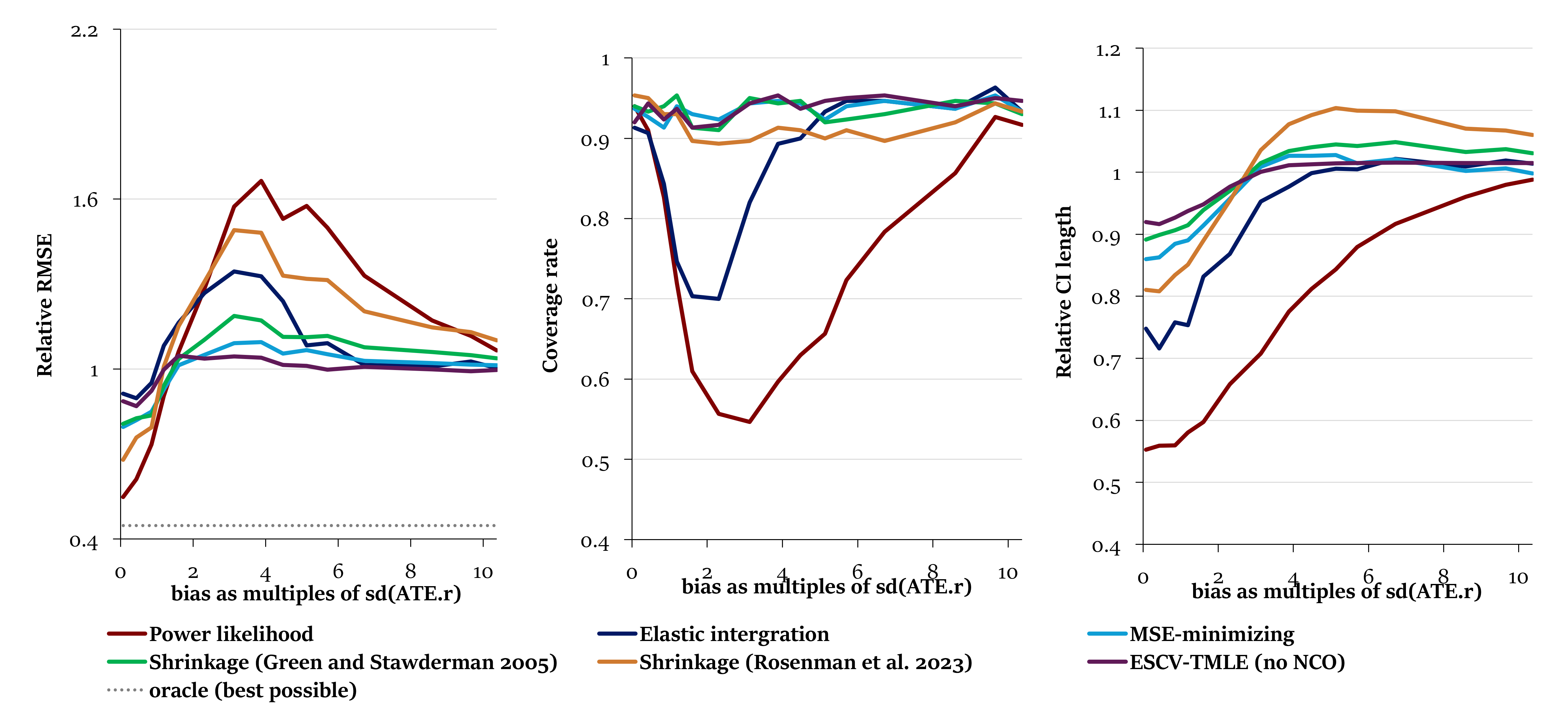}
    \caption{Relative RMSE, coverage rate and relative CI length of estimators that exhibit a trade-off \citep{green2005improved, rosenman2023combining, Oberst2022,lin2023many,dang2022cross,yang2023elastic}.}
    \label{fig:cont_tradeoff}
\end{figure}

Figure \ref{fig:cont_tradeoff} focuses on estimators that exhibit a trade-off pattern, showing a reduction in RMSE in some scenarios and an increase in others. Specifically, the leftmost panel of Figure \ref{fig:cont_tradeoff} demonstrates that when the RWD bias is small, typically less than twice the standard deviation of $\hat \tau_r$, the reduction in variance outweighs the introduced bias leading to an reduction in RMSE. However, as $\delta$ exceeds two standard deviations, the RMSE  increases compared to not incorporating RWD. This is primarily due to the bias overwhelming the benefit of reduced variance. As the RWD bias continues to increase, most methods gradually converge to the reference line, indicating that these methods recognize the extremely large RWD bias and respond by down-weighting the influence of the RWD to zero, relying solely on the RCT for estimation. 

The middle and the rightmost panels of Figure \ref{fig:cont_tradeoff} provide further details about the trade-off. Since the trade-off in RMSE involves a trade-off between squared bias and variance, it manifests in the coverage rate of the 95\% confidence interval and the relative length of these intervals. Additionally, there is a general trade-off between best-case reward and worst-case risk. Methods with larger potential efficiency gains are also more susceptible to higher worse-case RMSE. This is evident in Figure \ref{fig:cont_tradeoff}, where the order of lines from high to low before crossing the reference line is exactly reversed afterwards. This implies that aiming for a larger efficiency gain involves taking on greater risks.

Beyond the presented simulations, we have conducted additional simulations with varying parameter values. The relative behaviour remains somewhat consistent across these simulations. The power likelihood \citep{lin2023many}, elastic integration \citep{yang2023elastic}, and the shrinkage estimator by \cite{rosenman2023combining} are among the ``ambitious" methods with substantial potential efficiency gains but lower coverage rate in certain ranges of RWD bias. In contrast, the MSE-minimizing and ESCV-TMLE (no NCO) estimators tend to be ```conservative"; they offer a smaller potential reduction in uncertainty but maintain a coverage rate closer to the nominal rate.  

In this simulation setting, when the RWD is unconfounded, the power likelihood estimator achieves considerable efficiency gains, nearly reducing the RMSE to the best possible level and shortening the CI length by almost 45\%. However, this comes with the risk of CI under-coverage, with the worst-case coverage rate dropping below 60\%. As discussed in Section \ref{sec:bdb}, unlike frequentist methods that directly target the inference of ATE, the power likelihood approach selects the learning factor $\eta$ based on optimizing the expected predictive density over the target population. This means $\eta$ is chosen so that the $\eta$-indexed joint posterior of the model parameters best predicts the `true' data. Since this involves a joint distribution, $\eta$ is influenced not only by the ATE, which is primarily a function of the mean outcomes, but also the other features of the outcome distribution. For example, if the conditional variance of $Y$ differs between the RCT and the RWD---which is likely given the highly controlled environment of the RCT versus the numerous unmeasured uncertainties in RWD---the power likelihood method accounts for this difference. In such scenarios, stronger unmeasured confounding potentially increases the noise in $Y$. Consequently, the power likelihood approach typically selects a smaller $\eta$, rendering a more cautious inclusion of information from the RWD. The simulations in \cite{lin2023many} illustrated a scenario where the conditional volatility of $Y$ intensifies with the strength of unmeasured confounding. Their results showed that in such cases, the power likelihood method achieves efficiency gains while maintaining a coverage rate close to the nominal.

Key to the MSE-minimizing estimator is the weight $\lambda$ that balances the RCT and RWD estimates. As discussed in Section \ref{sec:weight_comb}, plug-in estimator $\hat \lambda$ underestimates the MSE-minimizing $\lambda$ due to Jensen's inequality. Consequently, less RWD is included than is theoretical optimal. This likely contributes to the estimator being more conservative compared to other methods.

\begin{figure}
    \centering
    \includegraphics[width=1\linewidth]{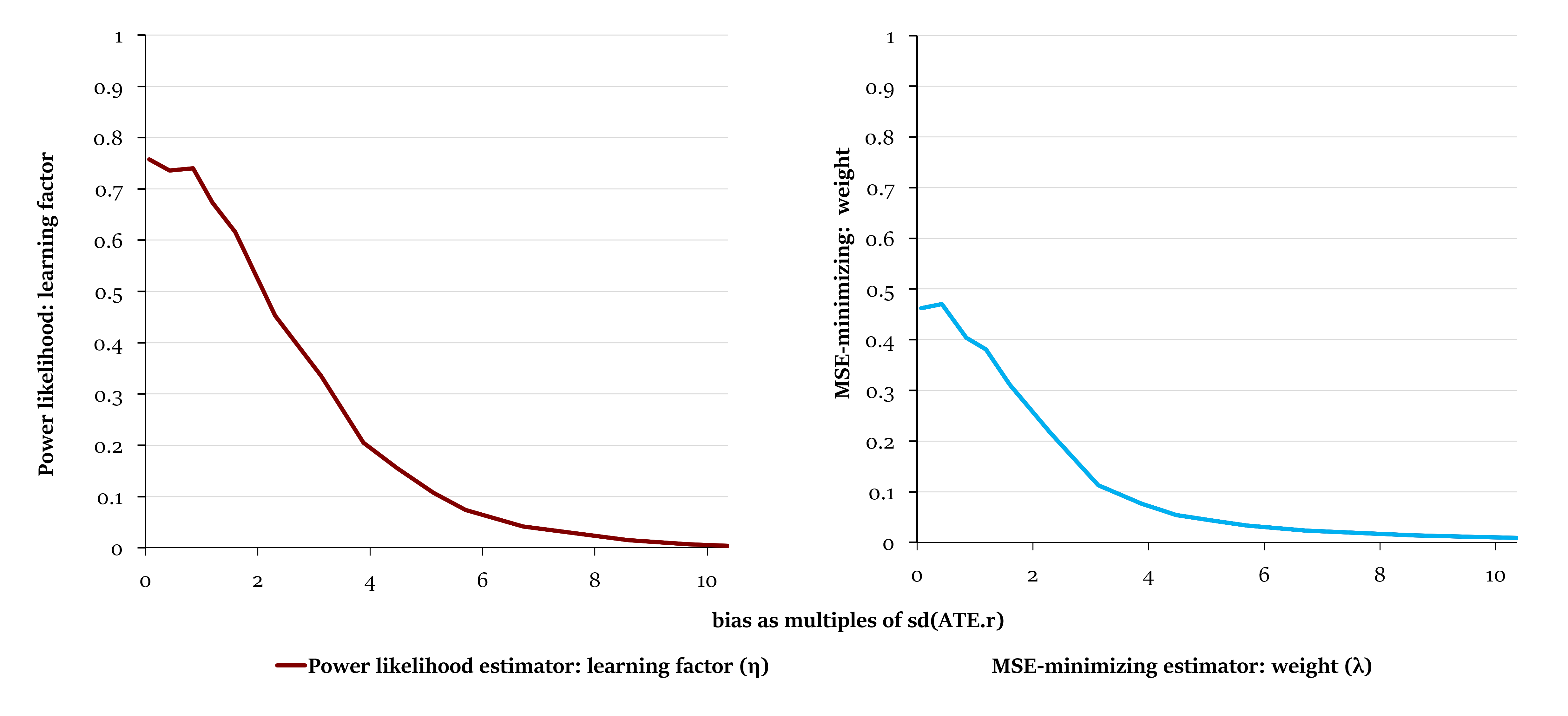}
    \caption{The average learning factor $\eta$ selected by power likelihood method \citep{lin2023many} and the average weight $\lambda$ assigned to $\hat \tau_o$ selected by the MSE-minimizing method \citep{Oberst2022}. Mote that due to differences in methodologies, the absolute levels of these two  parameters are not directly comparable.}
    \label{fig:learn_params}
\end{figure}

The learning factor $\eta$ in the power likelihood estimator and the weight $\lambda$ in the MSE-minimizing estimators play crucial roles in moderating the influence from the RWD. Both methods proposed ways to robustly select these parameters. Indeed, Figure \ref{fig:learn_params} demonstrates that both methods effectively downweight the RWD in response to increased bias. 

Regarding the shrinkage estimators in \cite{green2005improved} and \cite{rosenman2023combining}, as discussed in Section \ref{sec:shrinkage}, these methods are designed to improve the efficiency of the vector of strata estimates rather than directly enhancing the ATE. To get an estimate of the target estimand $\tau$, we introduce an extra step to multiply by a weight vector: 
\begin{align*}
    \hat \tau_{\operatorname{ATE},S1} &= \bs w^{T} \hat {\bs \tau}_{S1} &
    \hat \tau_{\operatorname{ATE},S2} &= \bs w^{T} \hat {\bs \tau}_{S2},
\end{align*}
where $\bs w^{T}$ is a $1 \times K$ vector with weights for each stratum in the target population.  We use $\hat w_k = n_{r,k}/n_r, k = 1,\dots K$ as an estimate. 

Theoretically, these shrinkage estimators should reduce the overall RMSE, where ``overall” refers to the sum, or a weighted sum, of strata estimator RMSEs. However, this theoretical guarantee of RMSE reduction does not necessarily extend to $\hat \tau_{\operatorname{ATE},S1}$ and $\hat \tau_{\operatorname{ATE},S2}$. This explains why, in certain ranges of $\delta$, neither the RMSE of the ATE is not consistently reduced, nor the length of CIs. Furthermore, as discussed in Section \ref{sec:shrinkage}, \cite{green2005improved} and \cite{rosenman2023combining} both suggested using the EB confidence interval for each entry of the vectors $\hat {\bs \tau}_{S1}$ and $\hat {\bs \tau}_{S2}$. However, since our focus is on $\hat \tau_{ATE,S1}$ and $\hat \tau_{ATE,S2} $, which are linear combinations of the strata shrinkage estimates, the point-wise CIs are not very useful. It is more appropriate to use the non-parametric bootstrap approach, which involves resampling each stratum in the RCT and RWD, respectively, and calculating the confidence intervals based on, in this implementation, 200 bootstrap samples.

\subsubsection{ESCV-TMLE \citep{dang2022cross}} \label{sec:both_escvtmle}
The ESCV-TMLE method offers the option to incorporate an NCO to enhance the estimation of bias. As discussed in Section \ref{sec:test-and-pool}, the NCO-integrated estimator relies on the U-compatibility assumption.  To investigate the sensitivity to this assumption, we modify the simulation parameterization by splitting the unmeasured confounder $U$ into two independent variables, $U_1 \sim \mathcal{N}(0,1)$ and $U_2 \sim \mathcal{N}(0,0.1)$, such that $U_1 + U_2$ is equivalent to $U$.
Specifically, 
\begin{align*}
    Y(a) \mid X_3,U_1,U_2 &\sim \mathcal{N}(0.5 + 0.2a + 0.1X_3 a + X_3 + U_1 + U_2,1) \\
    A_{obs} &\sim \bernoulli(\expit(-0.5 + X_1 + X_2 + X_3 + \psi U_1 + \psi U_2)).
\end{align*}
We further include three NCO variables:
\begin{align*}
    N_1 &\sim \mathcal{N}(1+ X_2+X_3 + U_1,2) &
    N_2 &\sim \mathcal{N}(1+ X_2+X_3 + U_2,2) &
    N_3 &\sim \mathcal{N}(0,2).
\end{align*}
In broad terms, $N_1$ captures $10/11$ of the total variation of unmeasured confounding, while $N_2$ captures the remaining $1/11$. We then refer to $N_1$ as a strong NCO and $N_2$ as a weak NCO. $N_3$ is simulated as a noise variable representing an extreme scenario where an NCO is entirely irrelevant.

\begin{figure}
    \centering
    \includegraphics[width=1\linewidth]{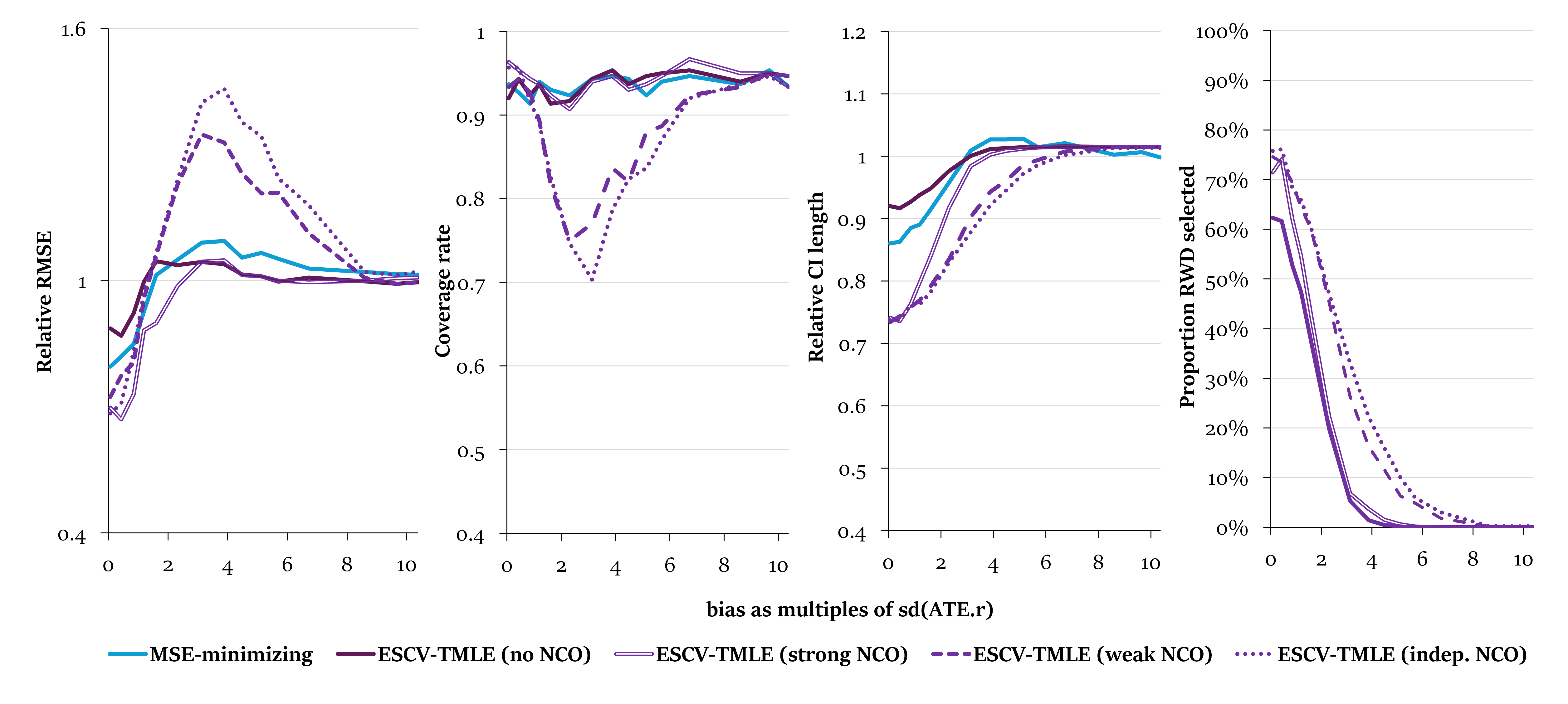}
    \caption{Relative RMSE, coverage rate, relative CI length, and average selected proportion of RWD of the ESVE-TMLE estimator \citep{dang2022cross} across scenarios without NCO, with a weak NCO, with a strong NCO and with an irrelevant NCO. The MSE-minimizing estimator \citep{Oberst2022} is included for reference.}
    \label{fig:cont_escvtmle}
\end{figure}

Figure \ref{fig:cont_escvtmle} compares the performance is different versions of the ESCV-TMLE estimator: without an NCO (solid line), with a weak NCO (dashed line), with a strong NCO (double line), and with an irrelevant NCO (dotted line). The MSE-minimizing estimator (blue solid line) is includes for reference. 
Interestingly, the inclusion of NCOs with varying degrees of relevance affects the estimator's behaviour differently. With a strong NCO, which captures most of the unobserved confounding, the performance is optimal. It shows substantial efficiency gains when RWD is unbiased and maintains a coverage rate consistently around 95\%. In this scenario, the ESCV-TMLE estimator dominates the MSE-minimizing estimator, as U-comparability is nearly satisfied the NCO strongly effectively enhances the bias estimation in the trade-off. However, if the NCO is weak or totally irrelevant, the estimator become overly ambitious, resulting in higher worst-case RMSE and CIs under-covering. In these cases, U-compatibility is violated, and the algorithm places undue confidence in an unreliable NCO.

The rightmost panel of Figure \ref{fig:cont_escvtmle} shows the average proportion of the RWD included, which helps explain the different behaviours across scenarios. Compared to not using an NCO, the versions with NCOs tend to select more RWD when the RWD bias is near zero, which is beneficial. However, the curves return back to zero more slowly than when not using an NCO. For the versions with a weak or irrelevant NCO, the proportions selected are consistently higher than when not using an NCO. This occurs because the squared-bias term in the selector function is underestimated under the illusion of minimal confounding bias due to the misleading NCO.

These findings suggest that when a candidate NCO is available, it is crucial to assess its relevance. A weak or irrelevant NCO can be misleading and may not function as intended. Therefore, using an NCO is not always better than not using one.

\subsection{Augmenting only the control arm of the RWD} \label{sec:control-only}

\subsubsection{PROCOVA \citep{schuler2020increasing}}

As described in Section \ref{sec:procova}, the PROCOVA method only uses the control arm of the RWD to fit a prognostic score model, $m(x) =  \E[Y \cmid A = 0, x, S=0]$, which is then added as a feature in the RCT dataset. Consequently, the interpretation of the x-axis in Figure \ref{fig:cont_base1} and \ref{fig:cont_cov1} is more nuanced. Since the PROCOVA method does not use $\hat\tau_o$, the impact of unobserved confounding is more on the expectation of the outcomes of the external controls . If there is no unobserved confounding, we expect $\E[Y \cmid A = 0, X, S=0] = \E[Y \cmid A = 0, X, S=1]$. The stronger the confounding, the larger the difference between these expectations.

In Figure \ref{fig:cont_base1} the PROCOVA line overlaps with the reference line, with a coverage rate constantly close to 95\% and no power gain, as shown in Figure \ref{fig:cont_cov1}. In this case, there is no benefit from data fusion. This is because the simulation setup, as outlined in Section \ref{sec:sim_setup}, assumes a linear outcome. As discussed in Section \ref{sec:procova}, the additional prognostic feature does not add any extra information if the outcome is linear, as noted by \cite{schuler2020increasing}. Additionally, the relative RMSE curve of the PROCOVA estimator is flat, indicating that the estimator remains unaffected by the difference between $\E[Y \cmid A = 0, X, S=0]$ and $\E[Y \cmid A = 0, X, S=1]$.

To understand how the PROCOVA estimator performs with non-linearity in the outcome, echoing the simulations in \cite{schuler2020increasing}, we introduce another standard normal variable $X_4$ and add a quadratic term $X_4^2$ to the outcome:
\begin{align*}
    Y(a) \mid X_3,U_1,U_2 &\sim \mathcal{N}\!\left(0.5 + 0.2a + 0.1a \times X_3  + X_3 + \alpha X_4^2 + U_1 + U_2, \, 1\right).
\end{align*}
\begin{figure}
    \centering
    \includegraphics[width=1\linewidth]{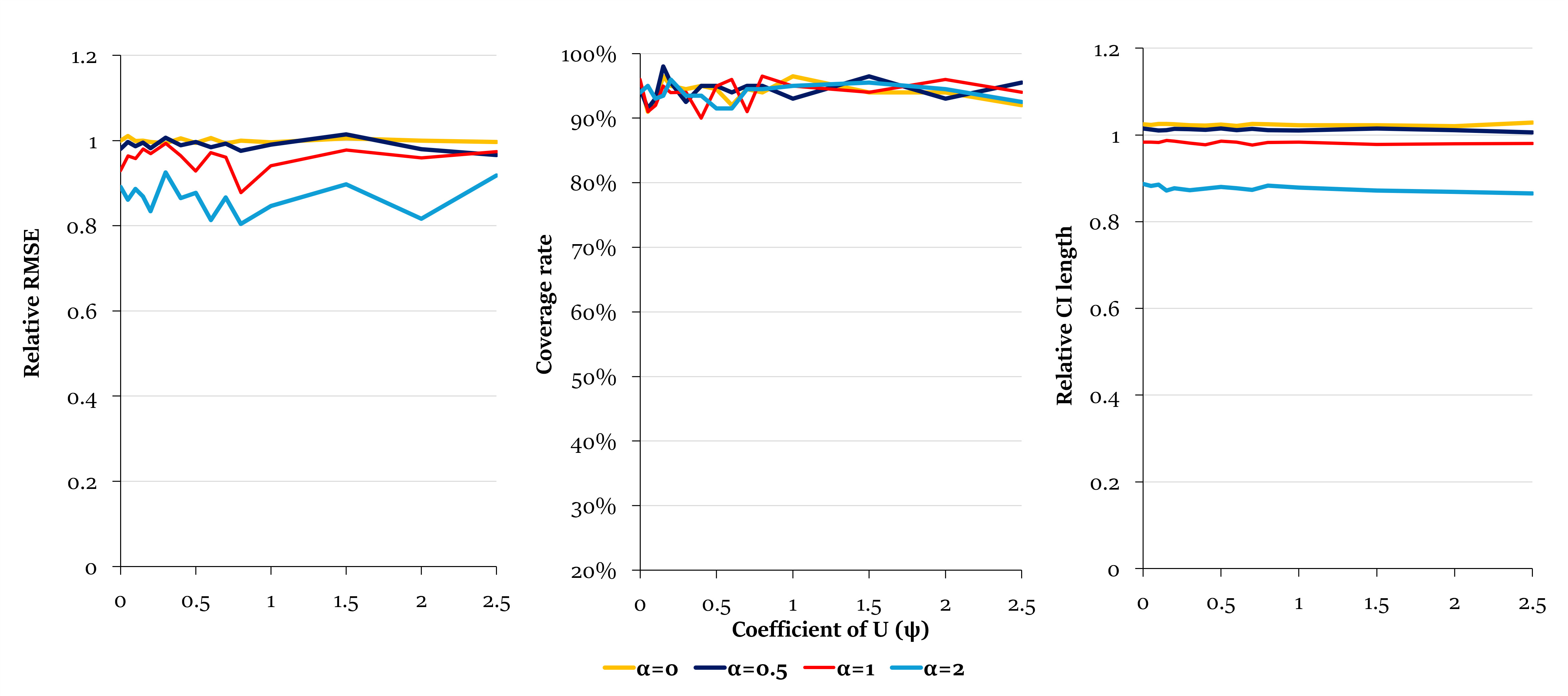}
    \caption{Relative RMSE, coverage rate and relative CI length of PROCOVA estimator \citep{schuler2020increasing} for outcomes with various degrees of curvature $\alpha = \{ 0,0.5,1,2\}$.}
    \label{fig:cont_procova}
\end{figure}
The coefficient $\alpha$ controls the degree of curvature in the outcomes. When $\alpha = 0$, the outcome is linear, as in the main simulation. Figure \ref{fig:cont_procova} shows that as $\alpha$ increases, the efficiency gain increases without compromising the coverage rate. Note that the rightmost panel of Figure \ref{fig:cont_procova} plots power on a relative scale, as the ratio against the power of $\hat \tau_r$. This is because, as $\alpha$ varies, the variance of $Y$ changes, making it sensible to compare on a relative rather than an absolute scale. In this specific simulation, when $\alpha = 2$, on average, the PROCOVA estimator shortens the CIs by approximately 12\%.

However, the above simulation assumes that the prognostic score is perfect, i.e. $m(X) = \E[Y(0)|X,S=1]$ using the notations in \cite{schuler2020increasing}. The authors relaxed this assumption in their Theorem 2, replacing it with the condition $\left|m(X)-\mu_0(X)\right| \xrightarrow{L_2} 0$, where $\mu_0(X) \equiv \E[Y(0)|X,S=1]$. To test the sensitivity to the perfect prognostic score assumption, we distinguished the curvature coefficient $\alpha$ for the RCT and RWD models by $\alpha_r$ and $\alpha_o$. Note that in this setting, the condition $\left|m(X)-\mu_0(X)\right| \xrightarrow{L_2} 0$ still holds. Table \ref{tab:procova} gives the relative RMSE in different ($\alpha_o$,$\alpha_r$) scenarios. We observe that the most efficiency gain is generally achieved when $\alpha_r = \alpha_o$, and the benefit diminishes when there is a discrepancy. Interestingly, when the RCT outcome is quadratic w.r.t.~$X_4$ but the RWD is linear without $X_4$, the RMSE of the PROCOVA estimator is even higher than $\hat \tau_r$. In this case, the prognostic score introduces more noise than it explains in $Y$'s residual variation. Essentially, in a most simplified multivariate regression scenario, the variance of the regression coefficient of the treatment $A$ can be expressed as:
\begin{align*}
     \Var( {\hat \beta_A} \cmid \textbf X) &= \frac{\sigma_{Y|X_1\dots X_d}^2}{n \times  \sigma^2_{A|X_1,\dots X_d}}\\
     & = \frac{\sigma_{Y|X_1\dots X_d}^2}{n_r\times  (1- R^2_{A|X_1,\dots X_d})\sigma_A^2},
\end{align*}
where $\sigma_{Y|X_1\dots X_d}^2$ is the residual variance of $Y$ after regressing on $X_1\dots X_d$ and $R^2_{A|X_1,\dots X_d}$ denotes the $R^2$ of regressing $A$ on all covariates. Although in an RCT, the treatment is randomized and $R^2_{A|X_1,\dots X_d}$ is expected to be zero in theory, this may not hold true in finite samples, especially in small RCTs. Adding an extra variable to the regression, regardless of its relevance, $\sigma_{Y|X_1\dots X_d}^2$ will either decrease or stay the same, and $R^2_{A|X_1,\dots X_d}$ will either increase or remain the same. If the additional variable fails to reduce the residual variance of $Y$, but instead increases $R^2_{A|X_1,\dots X_d}$, then variance of $\hat \beta_A$ will be higher as a result. This explains why, when $\alpha_r >0$ but $\alpha_o = 0$, the relative RMSE exceeds 100\%.


\begin{table}[h]
\centering
\begin{tabular}{c|cccc}
\toprule
 \textbf{Relative RMSE} & \textbf{$\alpha_r = 0$} & \textbf{$\alpha_r = 0.5$} & \textbf{$\alpha_r = 1$} & \textbf{$\alpha_r = 2$} \\
\midrule
\textbf{$\alpha_o = 0$} & 100\% & 112\% & 130\% & 177\% \\
\textbf{$\alpha_o = 0.5$} & 100\% & 98\% & 96\% & 89\% \\
\textbf{$\alpha_o = 1$} & 100\% & 100\% & 96\% & 88\% \\
\textbf{$\alpha_o = 2$} & 100\% & 100\% & 92\% & 84\% \\
\bottomrule
\end{tabular}
\caption{Relative RMSE of PROCOVA estimator when the prognostic score is imperfect.}
\label{tab:procova}
\end{table}

The PROCOVA method proposes a way to calculate the expected reduction in required RCT sample size prior to conducting the trial. The calculation is based on the expected correlation between the prognostic score and the residuals of $Y$ in the RCT. The estimation of the correlation relies entirely on the RWD and is sensitive to the assumption that the prognostic score is perfect. For example, when $\alpha_o = 2$ the estimated correlation is approximately 0.22, and one would conclude a roughly 4\% reduction in the required RCT sample size. However, if $\alpha_r = 0.5$, the actual correlation between the prognostic score and the RCT outcome residuals is close to 0, indicating that the utility of adding a prognostic score is overestimated. If we actually have reduced the RCT size by 4\% as suggested by the the procedure, we may end with inadequate power. This example highlights the limitation of the proposed sample size calculation procedure.

In the above discussion, we explored the sensitivity of the PROCOVA estimator to the degree of outcome non-linearity and the assumption of perfect prognostic score. Nonetheless, it maintains the nominal coverage rate regardless of the magnitude of confounding, safeguarding against Type I error, which is crucial for informing high-stakes decisions. The ability to calculate the expected sample size reduction also makes it a practical choice among data fusion methods.


\subsubsection{Power likelihood \citep{lin2023many}}
\begin{figure}
    \centering
    \includegraphics[width=1\linewidth]{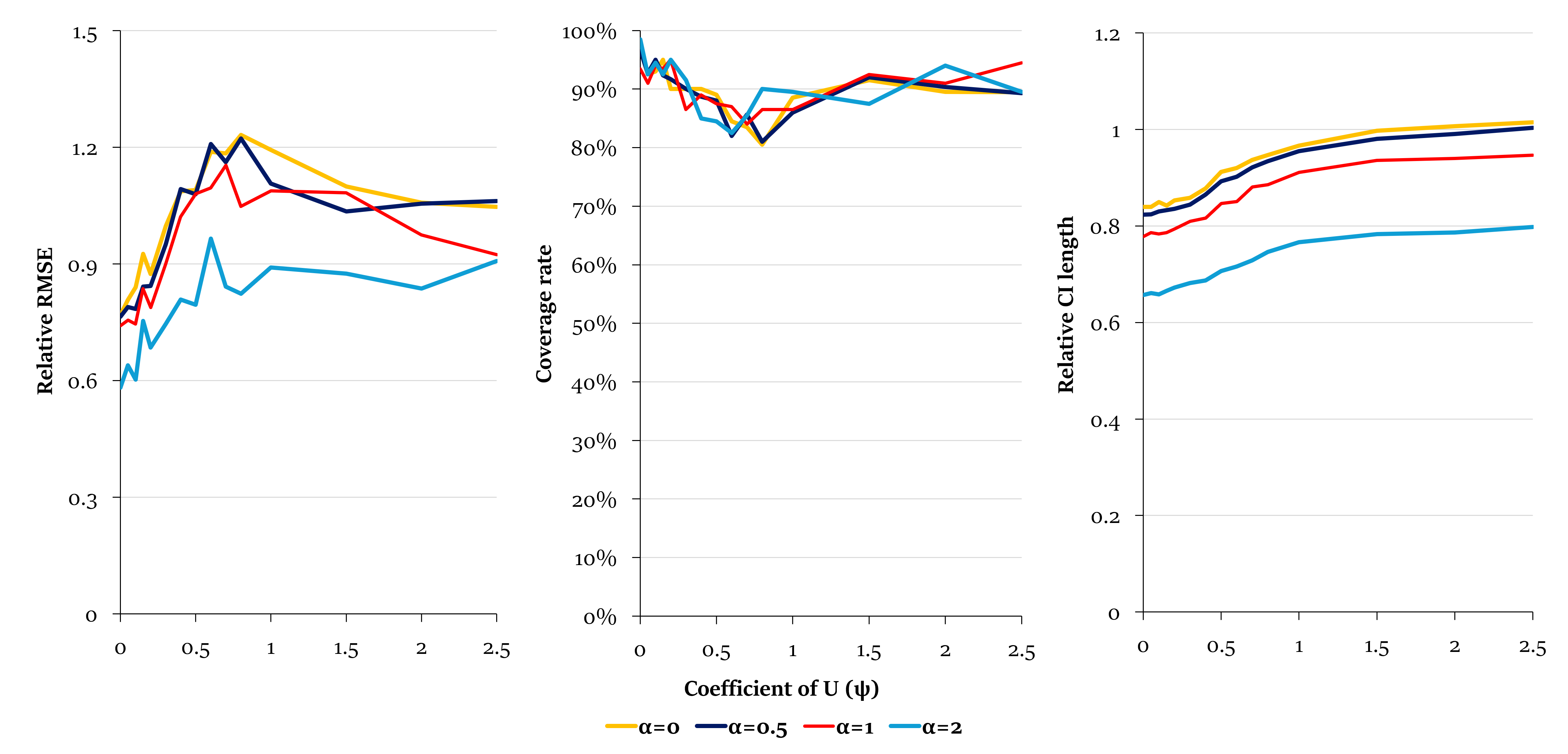}
    \caption{Relative RMSE, coverage rate and relative CI length of the power likelihood  estimator \citep{lin2023many} for outcomes with various degrees of curvature $\alpha = \{ 0,0.5,1,2\}$.}
    \label{fig:control_only_powlik}
\end{figure}

Figure \ref{fig:control_only_powlik} presents the performance of the power likelihood estimator.  The pattern is consistent with its application in integrating both arms: when confounding bias is small, the estimator effectively reduces variance while maintaining a coverage rate close to the nominal. However, as RWD bias intensifies, the estimator's RMSE exceeds that of $\hat \tau_r$ due to the excessive bias introduced by including RWD, resulting in CI undercoverage. As confounding bias continues to increase, making the RWD increasingly dissimilar to the RCT, the selected $\hat \eta$  to optimize ELPD is reduced to zero. This avoids extreme bias in the estimate, causing the RMSE to return to the reference line.

Interestingly, the reduction in variance compared to $\hat \tau_r$ is more pronounced as $\alpha$ grows. This might be due to the fact that with stronger non-linearity or more complex outcome models, the correctly-specified parametric outcome model outperforms non-parametric machine learning methods within the AIPW estimator, especially with small sample size. However, potential model misspecification in practice must be carefully considered.

\subsubsection{ESCV-TMLE \citep{dang2022cross}}

\begin{figure}
    \centering
    \includegraphics[width=1\linewidth]{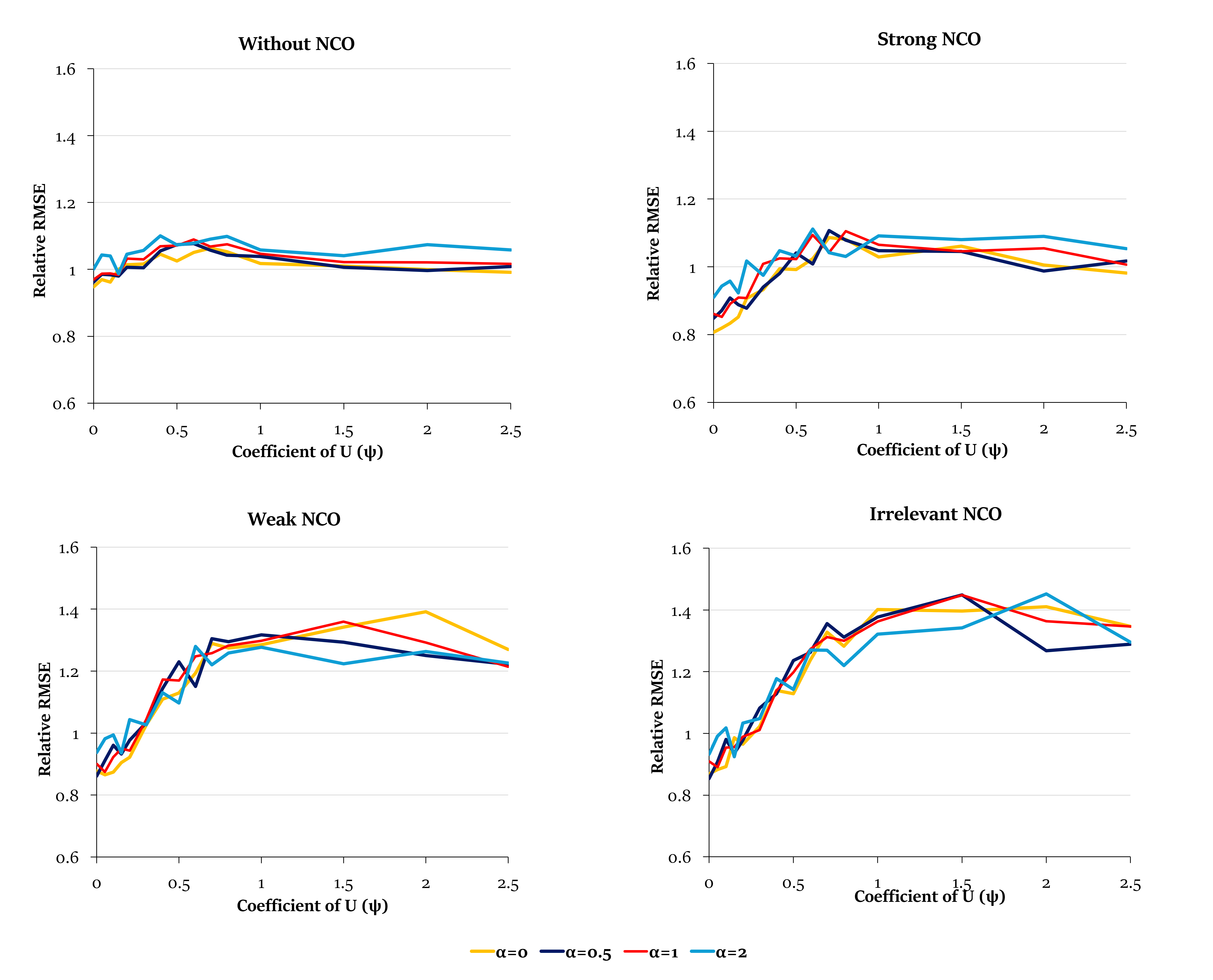}
    \caption{Relative RMSE, coverage rate and relative CI length of different versions of the ESCV-TMLE estimator \citep{dang2022cross}: without NCO, with a strong NCO, with a weak NCO and with an irrelevant NCO for outcomes with various degrees of curvature $\alpha = \{ 0,0.5,1,2\}$.}
    \label{fig:control_only_escvtmle}
\end{figure}

Figure \ref{fig:control_only_escvtmle} compares different versions of the ESCV-TMLE estimator. The findings mostly align with those in Section \ref{sec:both_escvtmle}. Including a weak or irrelevant NCO can expose the estimation to higher risk of bias. 

Notably, compared to PROCOVA and power likelihood estimators, the benefits of the ESCV-TMLE (no NCO) estimator are limited, especially as the outcome curvature intensifies. In the scenario where $\alpha = 2$, there is no reduction in the RMSE, and in fact, the ESCV-TMLE estimator performs worse than the RCT-only estimator, exhibiting a higher RMSE. \cite{van2024adaptive} reached a similar conclusion in their simulations. In their scenario (c) and (d), where complex non-linear terms were introduced into the generative model, the ESCV-TMLE failed to achieve any efficiency gain. They suspected that this might be due to the ESCV-TMLE not accurately estimating the bias structure or its cross-validation method not fully leveraging the data.

\section{Summary}\label{sec:discuss}
In this paper, we reviewed recent data fusion methods aimed at improving efficiency.

We first identified key challenges in data fusion, particularly the potential violations of comparability (Assumption \ref{assump:comparable}) and exchangeability of the RWD (Assumption \ref{assump:rwd_exchangeability}). These violations result in the ATE estimated from the RWD biased from the target ATE. When evaluating a data fusion method, it is crucial to examine these two assumptions and assess their plausibility in the specific analysis. Additionally, it is important to consider how the method adapts to such biases.

Secondly, while methodological developments typically provide a theoretical proof of asymptotic consistency and efficiency gains, the simulations in Section \ref{sec:sims} reveal important common patterns in their finite-sample behaviour.

Thirdly, we discussed the trade-offs between bias and efficiency. There is a notable trade-off along the scale of bias due to the violation of comparability and exchangeability assumptions. Potential efficiency gains when RWD bias is minimal come with the risk of excessive bias in the fusion estimate within the ‘danger zone’ (approximately 2--5 standard deviations of $\hat \tau_r$). This trade-off impacts the coverage rate and relative CI length, where methods capable of higher efficiency gains also risk producing estimates that are more biased.

Choosing the appropriate method is critical. Different methods inherently follow different trade-offs, and no single method dominates all others. What we can say with certainty is that while bias correction methods \citep{Kallus2018,yang2020improved} enhance the efficiency of HTE function estimation, these efficiency gains vanish when the HTE function is marginalized over the RCT sample to estimate the ATE. 
Additionally, the anchored thresholding method \citep{Chen2021} is not recommended due to its lack of robustness against bias from violations of Assumptions \ref{assump:comparable} and \ref{assump:rwd_exchangeability}. As bias increases, this method fails to revert to using the RCT alone.

Lastly, we explored the conservatism versus ambition in fusion methods. A more conservative fusion method is not necessarily better than an ambitious one as conservative methods often yield less efficiency gain even in optimal scenarios without confounding bias. If a method does not have the potential to deliver sufficient power gains in such scenarios, the data fusion analysis will not fulfil its intended purpose.

\begin{figure}
    \centering
    \includegraphics[scale = 0.5]{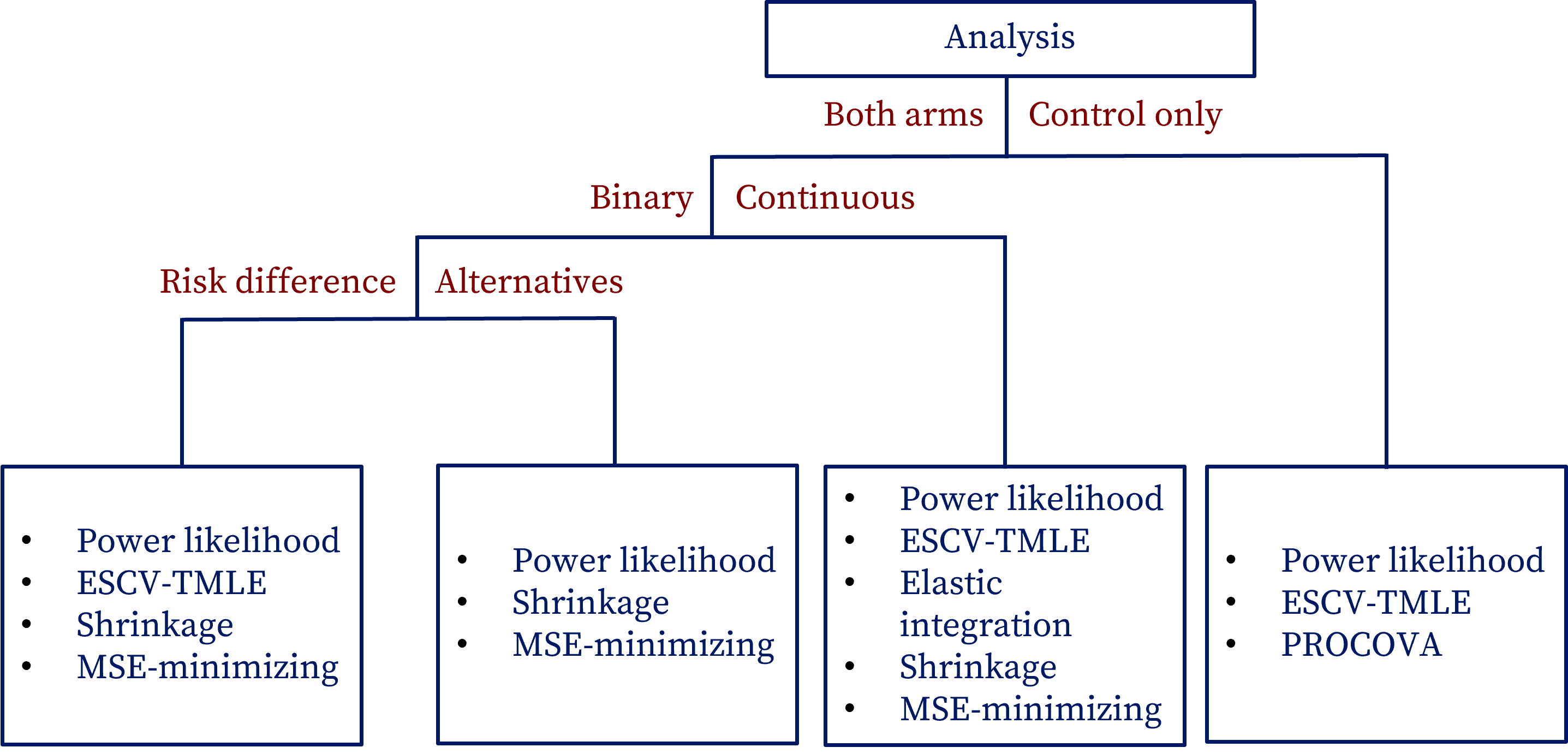}
    \caption{Decision framework for selecting appropriate data fusion methodologies based on analysis type, outcome type, and estimand in clinical research.}
    \label{fig:method_tree}
\end{figure}

Table \ref{tab:summ} outlines practical considerations when choosing a data fusion method. Figure \ref{fig:method_tree} provides a guide for selecting a method based on specific research needs.  A key initial consideration is the availability of treated subjects in RWD and whether to include them in the fusion. This is crucial when new treatments or policies are not yet available in the real world. If no unit received treatment in the RWD, the methods to consider are power likelihood \citep{lin2023many}, ESCV-TMLE \citep{dang2022cross} and PROCOVA \citep{schuler2020increasing}. Notably, PROCOVA provides Type 1 error protection and sample size calculation, although its power gains are often limited.

If including both arms from the RWD, the type of outcome dictates the next step. For continuous outcomes where the typical estimand is the difference in outcomes, options include power likelihood  \citep{lin2023many}, ESCV-TMLE \citep{dang2022cross}, elastic integration \cite{yang2023elastic}, shrinkage methods \citep{green2005improved,rosenman2023combining}, and MSE-Minimizing approach \citep{Oberst2022}. The choice among these should consider the trade-off between efficiency and bias, as well as implementation complexity.

For binary outcomes, the choice of method depends on the estimand. While the current implementation of ESCV-TMLE \citep{dang2022cross} is based on the risk difference, other methods including the power likelihood \citep{lin2023many}, the two shrinkage methods \citep{green2005improved,rosenman2023combining}, and the MSE-Minimizing approach \citep{Oberst2022} can adapt to alternative estimands like relative risk.

Through this paper, our goal is to provide researchers with a reference point and guidance for navigating various data fusion methods, enabling informed choices suited to specific research questions. For statisticians interested in further advancing this topic, we hope this paper presents a comprehensive overview of the current landscape, highlighting the challenges and opportunities present. In a subsequent paper, we apply the data fusion methods compared here to a real-world study, augmenting a medical trial for diabetic medication with US health claims data. This follow-up will detail practical considerations such as target trial emulation, handling covariate and outcome missingness, and variable selection.

\section*{Acknowledgments}
This work was conducted during Xi Lin's internship at Novo Nordisk, Denmark, in June to September 2023. We are grateful to Kajsa Kvist, Sky Qiu, Henrik Ravn, Kim Clemmensen, Morten Medici and the Joint Initiative of Causal Inference (JICI) for their valuable input and helpful discussions that contributed to this research. We also thank Chenyin Gao for his assistance in implementing the Elastic Integration method.

\appendix

\section{Implementation details of methods in simulation studies}\label{sec:implementation}
In this appendix, we provide a detailed description of the implementation of the data fusion methods used in our simulation studies presented in Section \ref{sec:sims}.

\subsection{Baseline estimators $\hat \tau_r$ and $\hat \tau_o$}
As outlined in Section \ref{sec:baseline_estimator}, we use AIPW estimators $\hat \tau_r$ and $\hat \tau_r$ as reference. The AIPW estimator involves a propensity model and an outcome model. For both, we employ a discrete \texttt{SuperLearner} \citep{van2007super} ensemble with a library consisting of a mean model and generalized linear models (GLM) (via R package \texttt{speedglm} \citep{enea2009fitting}). Given the simple generative models for $A$ and $Y$, this library should suffice. For simulations in Section \ref{sec:control-only} where a quadratic term is added, we also include generalized additive models \citep{hastie1987generalized} and random forest (via R package \texttt{ranger} \citep{wright2017ranger}) into the library to allow for non-linear flexibility. Additionally, we implemented a bound of $0.025$ on the propensity score to avoid extreme weights.

We use an empirical sandwich estimation of the variance of $\hat \tau_r$ following:
$$
\hat \sigma_r^2 =\frac{1}{n_r^2} \sum_{{i\in \mathcal{R}}}^{} \hat{I}_i^2
$$ where 
$$
\begin{aligned}
\hat{I}_i= & \left[ \frac{A_i \left (Y_i - \hat m_{1,r}(X_i)\right)}{\hat e_r(X_i)} - \frac{(1-A_i)\left(Y_i - \hat m_{0,r}(X_i) \right)}{1 - \hat e_r(X_i)} + \left (\hat m_{1,r}(X_i) - \hat m_{0,r}(X_i)\right ) \right] - \hat \tau_r.
\end{aligned}
$$ and similarly for $\hat\sigma_o^2$, estimated variance of $\hat \tau_o$.

\subsection{Shrinkage estimator \citep{green2005improved,rosenman2023combining}}

The shrinkage estimators in \cite{green2005improved} and \cite{rosenman2023combining} require the dimension of the vectors, i.e.~ $K$, to be greater than two and the elements to be independent. This necessitates stratifying the data. \cite{rosenman2023combining} discussed the performance of shrinkage estimators with different numbers of strata based on empirical evidence from simulations. However, their findings were inconclusive: the relative performance of the shrinkage estimators varied with different numbers of strata, and no single estimator consistently outperformed the others. The authors identified this as a potential topic for future research.

Despite the lack of definitive guidance, \cite{rosenman2023combining} found $\hat \tau_{S2}$ performed well with relatively few strata (2, 3, or 5). Based on this, we decided to stratify the RCT and RWD, respectively into four subgroups according to the values of $X_1$ and $X_3$. As outlined in Section \ref{sec:sim_setup}, $X_1$ and $X_3$ are binary variables, and the ATE estimates within these subgroups are independent, meeting the required assumptions.

To construct the elements in the shrinkage estimators $\hat {\bs \tau}_{S1}$ and $\hat {\bs \tau}_{S2}$, we use the AIPW algorithm to get the subgroups ATE estimates of the RCT, forming $\hat{\boldsymbol{\tau}}_{\boldsymbol{r}}$ vector. We apply the same procedure to the RWD. To estimate matrix $\bs\Sigma_r$, we use the sandwich estimators of variance of $\tau_{rk}$'s. For $\hat {\bs \tau}_{S1}$, there is a hyper-parameter $a$ and we set $ a = K-2$, which is a customary selection for James-Stein style estimators.

As mentioned in Section \ref{sec:sims_tradoff}, \cite{green2005improved} and \cite{rosenman2023combining} both proposed EB CIs for the elements of $\hat {\bs \tau}_{S1}$ and $\hat {\bs \tau}_{S2}$. However, since we are focused on the ATE, these element-wise CIs are not helpful. Therefore we used non-parametric bootstrap method, separately resampling the RCT and RWD data respectively, and obtained the quantiles from 200 iterations.

\subsection{Bias-correction methods}

\subsubsection{Experiment grounding \citep{Kallus2018}}
We follow the experimental grounding approach as described in Algorithm 1 in \cite{Kallus2018}. Specifically, we implement the following steps:
\begin{enumerate}[Step 1:]
    \item We fit two linear outcome regression models on the observational data: $m_1(X_1,X_2,X_3) = \E[Y \cmid X_1, X_2, X_3, A = 1,S=0]$ and $m_0(X_1,X_2,xX3) = \E[Y \cmid X_1, X_2, X_3,A= 0, S=0]$ , and calculate their difference as an estimate of CATE estimate $\hat \omega (\bs z, \bs c) = \hat m_1 (X_1,X_2,X_3) - \hat m_0 (X_1,X_2,X_3)$.
    \item We learn the bias correction function, assuming linearity,  $\xi(\bs x) = \beta^T \bs x$ where $\bs x$ is the vector of  $(x_1,x_2,X_3)^T$ and 
    $$
    \hat{\beta}=\underset{\beta}{\arg \min } \sum_{i \in \mathcal{R}}^{}\left(q_i Y_i-\hat{\omega}\left(\bs X_i\right)-\beta^{T} \bs X_i\right)^2,
    $$
    where $q_i = 2 \text{ if } T = 1$ and $q_i = -2 \text{ if } T = 0$. That is, we fit $\eta(\bs z, \bs c)$ through least square regression on $q_i Y_i-\hat{\omega}(\bs V_i)$ using the randomized data.
    \item We then calculate the ATE estimates as 
    $$\hat {\tau}_{XG} = \frac{1}{n_{r} }\sum_{i \in \mathcal{R}} \hat{\omega}(\bs X_i) + \hat \beta^T \bs {X}_i .$$ 
\end{enumerate}

For inference, we create confidence intervals using non-parametric bootstrap approach by resampling the RCT and RWD data respectively, obtaining the quantiles from 200 iterations.

\subsubsection{Confounding function \citep{yang2020improved}}
We implement the confounding function estimator using the R package \texttt{IntegrativeHTEcf} \citep{HTEcf2020}. This package assumes a linear form for both the confounding and HTE functions,  limiting its application to binary outcomes. Nonetheless, the linearity assumption is appropriate for the simulations presented. We specify the covariate set as $(X_1, X_2, X_3)$ and the HTE covariate set to be $X_3$, aligning with the correct data generation model.

The main function, \texttt{IntHTEcf}, outputs the estimated linear coefficients of the HTE and confounding functions, with variance estimated via an internal non-parametric bootstrap procedure. Since we are interested in the ATE, instead of the uncertainty around the HTE coefficients, we conduct our own bootstrap procedure outside the function to estimate the uncertainty of the ATE estimates.

\subsection{Adaptive power likelihood \citep{lin2023many}}
The implementation of this method draws upon R packages \texttt{causl} \citep{causl} and \texttt{ManyData} \citep{ManyData}. We follow these steps:

First, following the frugal parameterization, we parameterize the causal model as:

\begin{align*}
    X_1 &\sim \mathcal{N}(\beta_1, \phi_{1})\\
    X_2 &\sim \bernoulli(\operatorname{expit(\beta_2)})\\
    Y \mid do(A),X_3 &\sim \mathcal{N}(\beta_4 + \beta_5 A + \beta_6 X_3 A +  \beta_7 X_3, \phi_4)
\end{align*}
and the dependence between $X_1$,$X_2$,$Y$ follow a multi-variate Gaussian copula with constant correlation coefficients $\rho_1$, $\rho_2$ and $\rho_3$. We denote the vector consisting of $\beta$, $\phi$ and $\rho$ parameters as $\bs \theta$ . We do not need to assume parametric models for $A$ and $X_3$ as we are interested in the distribution conditional on treatment and the effect modifier.

Using the \texttt{fitCausl} function in \texttt{causl}, we  find $\hat {\bs \theta}_r$ and $ \hat {\bs \theta}_o$, which are the MLEs in the RCT and RWD. Next, we perform a grid search for the optimal learning rate $\eta \in [0,1]$. For each value of $\eta$ on the grid, we approximate the $\eta$-indexed posterior of $\bs \theta$ as described in Section 4.3 of \cite{lin2023many}. We then draw random samples from this posterior distribution, $\bs \theta^{(i)}$, for $i = 0,1,\dots, 2000$. For each sample $\bs \theta^{(i)}$, we approximate the ELPD using Leave-One-Out cross validation, implemented via the \texttt{loo} function in the R package \texttt{loo} \citep{vehtari2020loo}. Looping through the grid of $\eta$, we select the optimal value $\hat \eta$ that maximizes the approximated ELPD.

We then obtain the distribution of the ATE estimate based on $\bs \theta^{(i)}$ samples:

$$
\hat \tau_{PL}^{(i)} = \frac{1}{n_r}\sum_{j \in \mathcal{R}} \hat \beta_{5}^{(i)} + \hat \beta_{6}^{(i)} \times X_{3j},
$$
for $i = 1,2,\dots, 2000$. Finally, we derive the estimator $\hat \tau_{PL} = \frac{1}{2000} \sum_{i=1}^{2000} \hat \tau_{PL}^{(i)}$, as well as the credible interval based on the quantiles.

\subsection{Test-then-pool methods}

\subsubsection{Elastic integration \citep{yang2023elastic}}

We implement the elastic integration estimator using the R package \texttt{ElasticIntegrative} \citep{HTEcf2020}. This package is use-friendly and straightforward to use. The main function is \texttt{elasticHTE} and we specify $(X_1, X_2, X_3)$ as the covariate set for the propensity score model and $X_3$ for the HTE function, aligning with the data generative model. \texttt{elasticHTE} implemented flexible modeling through \texttt{SuperLearner} \citep{van2007super} and we specify the same libraries used for the AIPW estimation.

For inference, as the estimator is not regular, the variance estimated through non-parametric bootstrap will be biased; this is formalized with a proof in the supplementary materials of their paper. Therefore, the authors implemented a permutation-based bias estimation within the main function. We use the variance output from the function, although we modified the function slightly to also output the covariance. As we need to marginalize it over the covariate distribution of the RCT, we need to account for the correlation between coefficient estimates.

\subsubsection{ESCV-TMLE \cite{dang2022cross}}

We implement the ESCV-TMLE estimator using the R package \texttt{EScvtmle} \citep{ESCVTMLE2023}. This package is use-friendly and straightforward to use. The main function \texttt{ES.cvtmle} provides flexibility for combining either both arms or only the control arm of the RWD, the inclusion of a NCO, as well as different distribution families for the outcome and the NCO. We specify these settings accordingly for the simulations in Section \ref{sec:both_arms} and \ref{sec:control-only}. For the integrated \texttt{SuperLearner} \citep{van2007super} ensemble, we choose the discrete version with the same libraries used for the AIPW estimation. We use five folds for cross-validation.

We use the variance output from \texttt{ES.cvtmle} to construct 95\% confidence intervals.

\subsection{Weighted combination methods}

\subsubsection{MSE-minimizing estimator \citep{Oberst2022}}
We implement the MSE-minimizing estimator following
\begin{align*}
    \hat{\tau}_{\hat{\lambda}}=\hat{\lambda} \hat{\tau}_o+(1-\hat{\lambda}) \hat{\tau}_r && \text{where} && \hat{\lambda}=\frac{\hat{\sigma}_r^2}{\left(\hat{\tau}_r-\hat{\tau}_o\right)^2+\hat{\sigma}_r^2+\hat{\sigma}_o^2}.
\end{align*}
As $\hat{\tau}_e$ and $\hat{\tau}_o$ are independent, the covariance term in the original formula of $\hat \lambda$ drops out. For inference, we generate confidence intervals by applying a non-parametric bootstrap method, separately resampling the RCT and RWD data and extracting quantiles from 200 iterations.

\subsubsection{Anchored-thresholding estimator \citep{Chen2021}}

We implement the anchored-thresholding estimator by first calculating the estimated bias $\hat \delta_\gamma$ through soft-shresholding:
$$
\hat{\delta}_\gamma= \begin{cases}\operatorname{sgn}(\hat{\delta})\left(|\hat{\delta}|-\gamma\cdot \sqrt{\hat{\sigma}_r^2+\hat{\sigma}_o^2}\right) & \text { if }|\hat{\delta}| \geq \gamma \cdot \sqrt{\hat{\sigma}_r^2 +\hat{\sigma}_o^2} \\ 0 & \text { otherwise. }\end{cases}
$$.

As discussed in Section \ref{sec:sims_tradoff}, hyper-parameter $\gamma$ is important in controlling the combination and the larger $\gamma$ is, the higher the potential bias is. \cite{Chen2021} state that $\gamma$ should be of asymptotic order $\sqrt{\log \left(\min(n_r,n_o))\right)}$, so we set $\gamma = \sqrt{\log 300}$. We then plug $\hat \delta_\gamma$ into the estimator:
$$
\hat \tau_{AT} = (1 - \hat \omega) \hat \tau_r + \hat \omega (\hat \tau_o + \hat \delta_\gamma),
$$ 
where $\hat \omega = \hat{\sigma}_r^2 / (\hat{\sigma}_r^2 + \hat{\sigma}_o^2)$. We use a non-parametric bootstrap to obtain the 95\% confidence interval.

\subsection{Prognostic adjustment \citep{schuler2020increasing}}

The implementation of the PROCOVA approach is relatively straight forward. We first empirically centre the covariates as well as the treatment indicator. Then employing the \texttt{SuperLearner} ensemble, we fit a prognostic score model and make predictions on the RCT, creating prognostic scores $\hat m(X)$.Then, we estimate the treatment effect $\hat \beta_A$ using a linear regression adjusted for the empirically centered covariates, prognostic score, and their interactions with the treatment.

We use the sandwich estimator of variance of the $\hat \beta_A$ to construct the 95\% confidence interval.

\bibliographystyle{abbrvnat}
\bibliography{refs}
\end{document}